\documentclass[aps,prb,showpacs,superscriptaddress,twocolumn,10pt,floatfix]{revtex4-1}
\usepackage{graphicx, color}
\usepackage[caption=false]{subfig}
\usepackage{amsmath, amsthm, amssymb}
\usepackage{bm}
\usepackage{natbib}
\usepackage{hyperref}

\newcommand{\mb}[1]{\mathbf{#1}}
\newcommand{\pd}{\partial}

\newcommand{\tr}{\textrm{Tr}}

\newcommand{\mbb}[1]{\mathbb{#1}}

\def\sgn{\textrm{sgn}}
\newcommand{\comments}[1]{} 

\begin{document}
\title{Josephson effect in superconducting wires supporting multiple Majorana edge states}
\author{Doru Sticlet}
\email{sticlet@pks.mpg.de}
\affiliation{Laboratoire de Physique des Solides, CNRS UMR-8502, Universit\'e Paris Sud, 91405 Orsay, France}
\author{Cristina Bena}
\affiliation{Laboratoire de Physique des Solides, CNRS UMR-8502, Universit\'e Paris Sud, 91405 Orsay, France}
\affiliation{Institute de Physique Th\'eorique, CEA/Saclay, Orme des Merisiers, 91190 Gif-sur-Yvette Cedex, France}
\author{P. Simon}
\affiliation{Laboratoire de Physique des Solides, CNRS UMR-8502, Universit\'e Paris Sud, 91405 Orsay, France}
\date{\today}
\begin{abstract}

We study superconducting-normal-superconducting (SNS) Josephson junctions in one-dimensional topological superconductors which support more than one Majorana end mode. 
The variation of the energy spectrum  with the superconducting phase is investigated by combining  numerical diagonalizations of tight-binding models describing the SNS junction together with an analysis of appropriate low-energy effective Hamiltonians. We show that the $4\pi$-periodicity characteristic of the fractional dc Josephson effect is preserved. Additionally,  the ideal conductance of a NS junction with a topological supraconductor, hosting two Majorana modes at its ends, is doubled compared to the  single Majorana case. Last, we illustrate  how a nonzero superconducting phase gradient can potentially destroy the phases supporting multiple Majorana end states.
\end{abstract}
\maketitle

\section{Introduction}
Majorana fermions have recently attracted considerable attention in the condensed-matter community due to their exotic character (see recent reviews, Refs.~\onlinecite{Alicea12,Beenakker12}).
The fact that they obey non-Abelian statistics makes  them interesting candidates for quantum computation.\cite{Nayak08}
Many systems hosting Majorana fermions have been proposed. One-dimensional semiconducting wires with a strong spin-orbit coupling in the proximity of an $s$-wave superconductor, and subject to a Zeeman field seem to offer a promising route to observe Majorana fermions. Recent experiments using  InSb\cite{Mourik12,Xu12} or InAs\cite{Das12} quantum wires have reported signatures consistent with Majorana physics in conductance measurements in superconducting-normal (SN) junctions, though it has been shown that the interpretation of the data may not be so straightforward because of the presence of disorder.\cite{Bagrets12,Liu12,Pientka12,Rainis12}

Significant signatures of Majorana physics have also been predicted when two superconducting quantum wires are brought into contact to form a SNS Josephson junction. It has first been shown by Kitaev\cite{Kitaev} for single-channel topological superconducting wires that the two Majorana end states across the junction couple to each other and generate  a fractional Josephson effect:  instead of the usual $2\pi$, the Josephson current exhibits a periodicity of $4\pi$ with the phase difference between the two superconductors. This doubling of the periodicity can be interpreted as the tunneling of ``half'' of a Cooper pair, hence the term fractional.  This prediction has later been extended to many different systems. \cite{Kwon03,Yakovenko04,Fu09,Badiane11,Ioselevich11,Alicea11,Jiang11,San-Jose12,Pikulin12,Platero12}
The fractional Josephson effect has been shown to be robust in a realistic system, even when the Majorana fermions are coupled to other fermionic end states.\cite{Law11} 
A direct consequence of the fractional Josephson effect is the presence of anomalous Shapiro steps at double the expected voltage; such doubling has recently been reported experimentally.\cite{Rokhinson}
This fractional Josephson effect can be understood by a simple low-energy effective Hamiltonian that couples the two Majorana fermions $\gamma_1$ and $\gamma_2$ across the one-dimensional (1D) junction, $H_{\rm eff}=i\Gamma(\phi) \gamma_1\gamma_2$, where $\Gamma$ is the coupling, $\phi$ is the superconducting phase difference, and  $\Gamma$ satisfies $\Gamma(\pi)=0$. The operator $i\gamma_1\gamma_2$ is obviously a conserved quantity of the Hamiltonian for all values of $\phi$. This guaranties a level crossing at $\phi=\pi$ in the evolution of the energy spectrum with the phase $\phi$. 

However, the situation is far from obvious if we consider 1D junctions between wires supporting several Majorana fermions. The question we are addressing in this work is whether the fractional Josephson effect survives in such situations, and whether the periodicity of the phase/current dependence is modified.

This question is relevant in the light of recent studies of a generalized  1D Kitaev chain which supports  
two end Majorana fermions at each extremity,\cite{Chakravarty} and of 1D topological superconducting wire that can support multiple end Majorana fermions. It has been recently shown by Tewari and Sau\cite{Tewari} that the effective low-energy Bogoliubov-de Gennes (BdG) Hamiltonian  currently used to describe narrow semiconducting wires with a strong spin-orbit coupling, in the presence of a Zeeman splitting and in the proximity of an $s$-wave superconductor\cite{sau10,oreg10} can be made purely real, thereby restoring  effectively some time-reversal symmetry (TRS). Consequently,
they belong to the symmetry class BDI in the classification of the {\it non-interacting} band Hamiltonians.\cite{Schnyder08,Schnyder09,Kitaev09} The BDI class implies a $\mathbb{Z}$ topological invariant instead of a $\mathbb{Z}_2$ invariant for the D symmetry class.\cite{*[{It was recently shown in }][{ that in a 1D interacting system the $\mathbb{Z}$ topological invariant reduces to  $\mathbb{Z}_4$}.] wen12} Thus, a 1D topological superconducting wire belonging to the BDI class can support multiple end Majorana fermions. 

Along with the time-reversal symmetry (TRS) and an intrinsic particle-hole (p-h) symmetry of the low-energy BdG Hamiltonians, there is a chiral symmetry\cite{Schnyder08,Schnyder09,Kitaev09} which prevents the pairs of Majorana zero modes to recombine at the edge. The chiral symmetry can be effectively realized in sufficiently narrow wires, typically for a wire width smaller than the spin-orbit coupling length.\cite{Diez12} In such system, it was shown that two proximate robust Majorana fermions can be created by manipulating the sign of spin-orbit coupling and also that they are robust against disorder and small variations of magnitude of magnetic fields.\cite{Ojanen12} 
The presence of several Majorana fermions at one edge
of the superconducting wire opens several Andreev transport channels in SN junctions and therefore the conductance can reach the value $2e^2/h\times Q$ with $Q\in\mathbb{Z}$.\cite{Diez12}

The outline of the paper is as follows: Sec.~\ref{sec:ham} presents the tight-binding model for a chain supporting multiple Majorana at its ends, and some general symmetry arguments. Section~\ref{sec:joes} analyzes the energy levels and a low-energy effective Hamiltonian in Josephson junctions between topological superconductors with one and two Majorana modes at their ends: the $1-1$ junction in~\ref{subsec:11}, the $2-2$ junction in~\ref{subsec:22} and the $1-2$ junction in~\ref{subsec:12}. The general case of the $p-q$ junction is briefly discussed in Sec. \ref{subsec:pq}. Then Sec.~\ref{sec:transp} obtains the conductance of an NS junction where the superconductor is topological and supports two Majorana modes at each end.  Section~\ref{sec:grad} is devoted to breaking the $\mbb Z$ phase through the addition of a uniform superconducting phase gradient. Section~\ref{sec:conc} presents the conclusions of the study.

\section{Model Hamiltonian and symmetry properties}
\label{sec:ham}
In this section, we consider the  1D model of spinless fermions, proposed  by Niu {\it et al.} in Ref.~[\onlinecite{Chakravarty}], which
supports several Majorana end states. The Hamiltonian reads 
\begin{eqnarray}\label{eq:H}
H&=&\sum_j\big[-\mu(1-2c^\dag_jc_j)-\lambda_1(c^\dag_jc_{j+1}
+c^\dag_jc^\dag_{j+1}+\text{H.c.})\notag\\
&&-\lambda_2(c^\dag_{j-1}c_{j+1}+c^\dag_{j-1}c^\dag_{j+1}
+\text{H.c})
\big],
\end{eqnarray}
where $\lambda_1$ corresponds simultaneously to the nearest-neighbor (NN) hopping amplitude and to the nearest-neighbor superconducting gap,
while $\lambda_2$ denotes the next-nearest-neighbor (NNN) hopping amplitude and next-nearest-neighbor superconducting gap. In what follows, the chemical potential $\mu$ is set to 1 and $\lambda_1$ is assumed positive.
When $\lambda_2=0$, this Hamiltonian in Eq.~(\ref{eq:H}) corresponds to the Kitaev model.\cite{Kitaev}
Before analyzing the properties of $H$, let us provide some general symmetry arguments which explain why a general 1D Hamiltonian can sustain phases with more than one Majorana end state.

\subsection{Symmetry arguments and the role played by distant site hopping}
A 1D superconducting system can have multiple Majorana bound states at its ends when the system is time-reversal invariant (TRI) and exhibits particle-hole symmetry.\cite{Chakravarty, Tewari} For the two-band Bogoliubov-de Gennes (BdG) Hamiltonian presented here, this can be seen from the following simple argument. A general two-band BdG Hamiltonian $H_{\rm BdG}$ obeys the particle-hole (p-h) symmetry by construction, and can be written in the p-h basis as
\begin{equation}
H_{\rm BdG}=\mb h\cdot\bm\tau,
\end{equation}
where $\tau$'s are the Pauli matrices in the p-h space. Note that the p-h symmetry requires that
\begin{align}
h_{1}(k)&=-h_{1}(-k),\notag\\
h_{2}(k)&=-h_{2}(-k),\notag\\
h_3(k)&=h_3(-k).
\end{align}
The time-reversal operator for spinless fermions is just the complex conjugation operator. If the system is TRI, the components of $\mb h$ obey the following constraints
\begin{align}
h_{1}(k)&=h_{1}(-k),\notag\\
h_2(k)&=-h_2(-k)\notag\\
h_{3}(k)&=h_{3}(-k).
\end{align}
Particle-hole and time-reversal symmetries impose the chiral symmetry represented by the operator $\tau_1$ which anti-commutes with the Hamiltonian
\begin{equation}
\{H_{\rm BdG},\tau_1\}=0.
\end{equation} 
If $H_{\rm BdG}$ obeys these symmetries, then it follows that $h_1$ must vanish.  Hence $H_{\rm BdG}$ has only two remaining components and  $\hat{\mb h}$ defines therefore a mapping from the Brillouin zone (BZ)  to the Bloch ``circle'', $\hat{\mb h}:T^1\to S^1$. The mapping is characterized by a winding number $w$ which is an integer.\cite{Tewari} Therefore a two-band BdG Hamiltonian belongs to the topological BDI class characterized by a ${\mathbb Z}$ topological invariant.\cite{Schnyder09,Kitaev09,Ryu10}

The expression of the winding number reads
\begin{equation}
w=\frac{1}{2\pi}\int_0^{2\pi}dk(\hat{\mb h}\times\pd_k\hat{\mb h})_1.
\end{equation}
The discrete form for the winding number reads
\begin{equation}\label{eq:winding}
w=-\frac{1}{2}\sum_{k\in\ker h_2}\sgn[h_3\pd_kh_2 ].
\end{equation}
Note from the symmetry constraints that the kernel of $h_2$ contains at least the special BZ points $0$ and $\pi$.
To create more Majorana bound states at one end, the winding number must have $|w|>1$. This implies that $\ker h_2$ must contain more points in addition to $0$ and $\pi$. 
This can happen for example by having more sites in the unit cell. Then new nodes in the energy dispersion develop for $k\in[0,\pi]$ and can lead to higher winding number. 

\subsection{Phase diagram}
\begin{figure}[t]
\centering
\includegraphics[width=6.5cm]{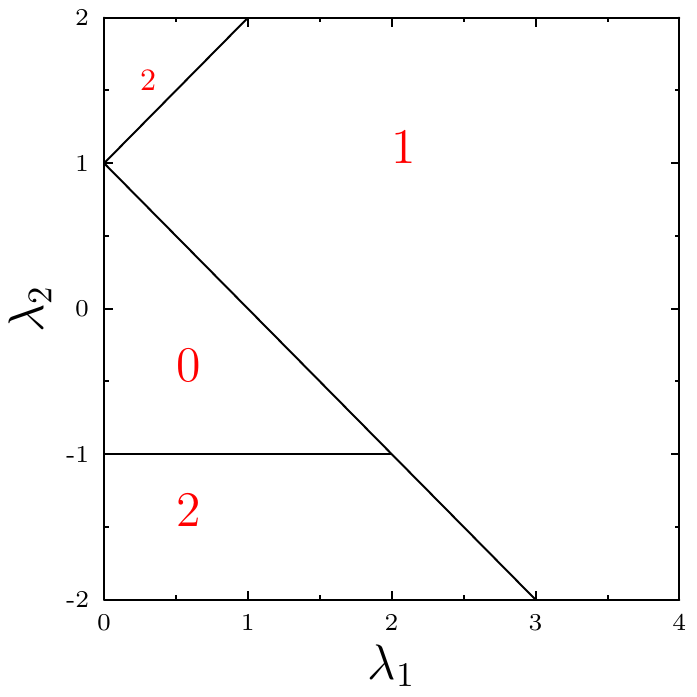}
\caption{(Color online.) Phase diagram for the Hamiltonian in Eq.~(\ref{eq:H}). The insulating phases are separated by black lines. Each insulating phase is characterized by a winding number represented in red.}
\label{fig:phdiag}
\end{figure}

The phase diagram of $H$ has been established in Ref.~\onlinecite{Chakravarty}. Here we recover this phase diagram in a different manner, by using  Eq.~(\ref{eq:winding}) to unambiguously characterize each topological phase in the $(\lambda_1,\lambda_2)$. This phase diagram is drawn for completeness in Fig.~\ref{fig:phdiag}.

The phase diagram associated with $H$ is characterized by phases with
$w=0,1,2$. Indeed, for $\lambda_2>1+\lambda_1$ or $\lambda_2<-1$ and $\lambda_2<1-\lambda_1$, $H$ can sustain a phase with two Majorana zero modes
localized at each end.\cite{Chakravarty}

\section{Josephson junctions}
\label{sec:joes}
In this section, we analyze the Josephson effect in the presence of several Majorana zero modes.
Let us consider short Josephson junctions between two wires which can sustain several Majorana end states. The model Hamiltonian reads
\begin{equation}
H=H_L+H_R+H_T
\end{equation}
The Hamiltonian for the left wire, $H_L$, is described by
the Hamiltonian $H$ in Eq.~(\ref{eq:H}) characterized by parameters $(\lambda_1^L,\lambda_2^L)$.
The right wire is characterized by $H_R$, which is obtained from $H_L$ by changing  ($c^\dag_ic^\dag_j\to c^\dag_ic^\dag_je^{i\phi}$) in Eq. (\ref{eq:H}). Note that the same phase variable phase $\phi$ is attached to the NN and NNN pairing terms. The second wire is characterized by the parameters $(\lambda_1^R,\lambda_2^R)$. The tunneling Hamiltonian for the short junction can be modeled as\normalsize
\begin{equation}
H_T
=-(\lambda_1^Lc_m^\dag c_{m+1}+\lambda_2^Lc_{m-1}^\dag c_{m+1}
+\lambda_2^Lc_{m}^\dag c_{m+2}+{\rm H.c.}),
\end{equation}
with $m$ the last site of the the left region and $m+1$ the first site in the right region. Choosing the hopping integral in the junction to be equal to the left region hopping $\lambda_2^L$ is purely conventional.

\begin{figure}[t]
\centering
\includegraphics[width=0.7\columnwidth]{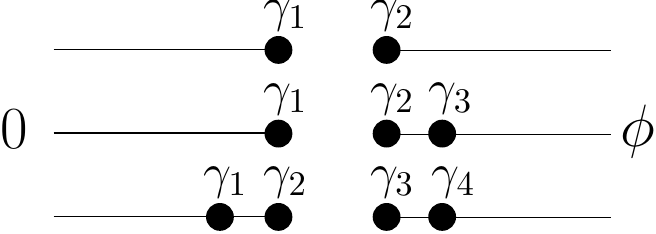}
\caption{To form Josephson junctions, wires characterized by winding numbers $1-1$, $1-2$ and $2-2$ respectively, are brought  into contact. Without loss of generality, the right-hand-side superconductors have real order parameters, while the left-hand-side superconductors have a superconducting (SC) phase $\phi$. The low-energy Hamiltonian is assumed to contain only phase-dependent coupling terms between the Majorana fermions.}
\label{fig:junctions}
\end{figure}

Each wire is labeled by a topological index $w^\alpha=0,1,2$ with $\alpha=L,R$. We have  analyzed various $w^L-w^R$ junctions which support Majorana fermions. We present below the $1-1$, $1-2$ and $2-2$ junctions as sketched in Fig.~\ref{fig:junctions}.

Before analyzing the Josephson effect in junctions made with wires supporting multiple Majorana fermions, one may ask if a complex superconducting order parameter in a topological superconducting wire can have an effect on its phase diagram. However, while $H_R(\phi)$ is generically complex, a uniform phase $\phi$ can be gauged away, yielding a real Hamiltonian and the same phase diagram as the one depicted in Fig.~\ref{fig:phdiag}. 

\subsection{The \texorpdfstring{$1-1$}{1-1} Josephson junction}
\label{subsec:11}
\begin{figure}[t]
\centering
\includegraphics[width=\columnwidth]{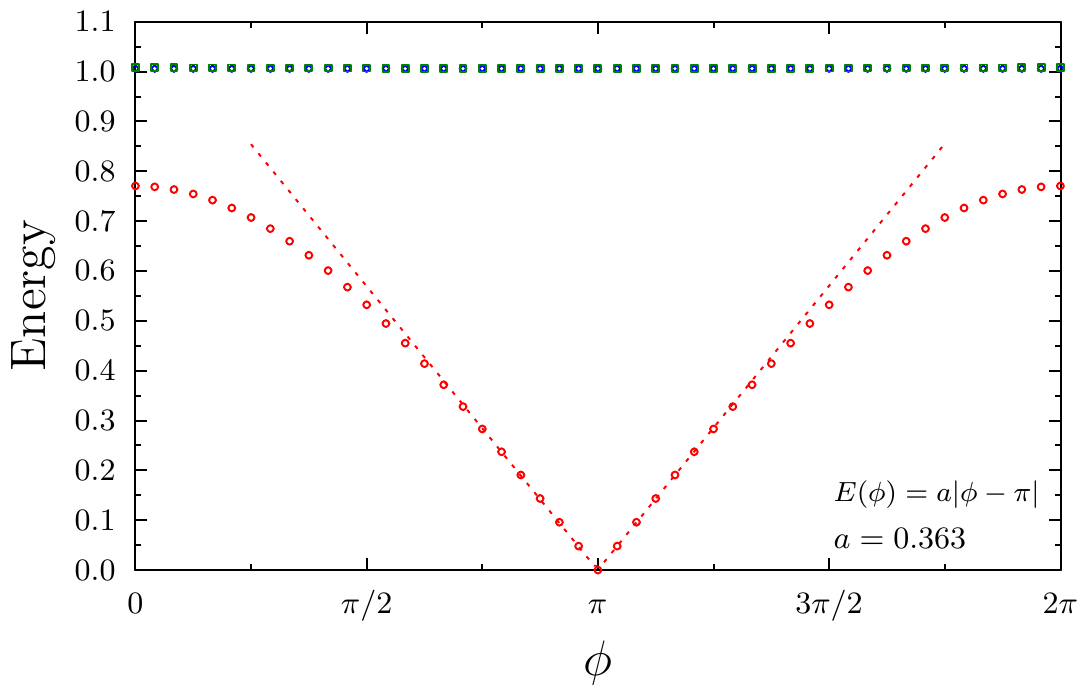}
\includegraphics[width=\columnwidth]{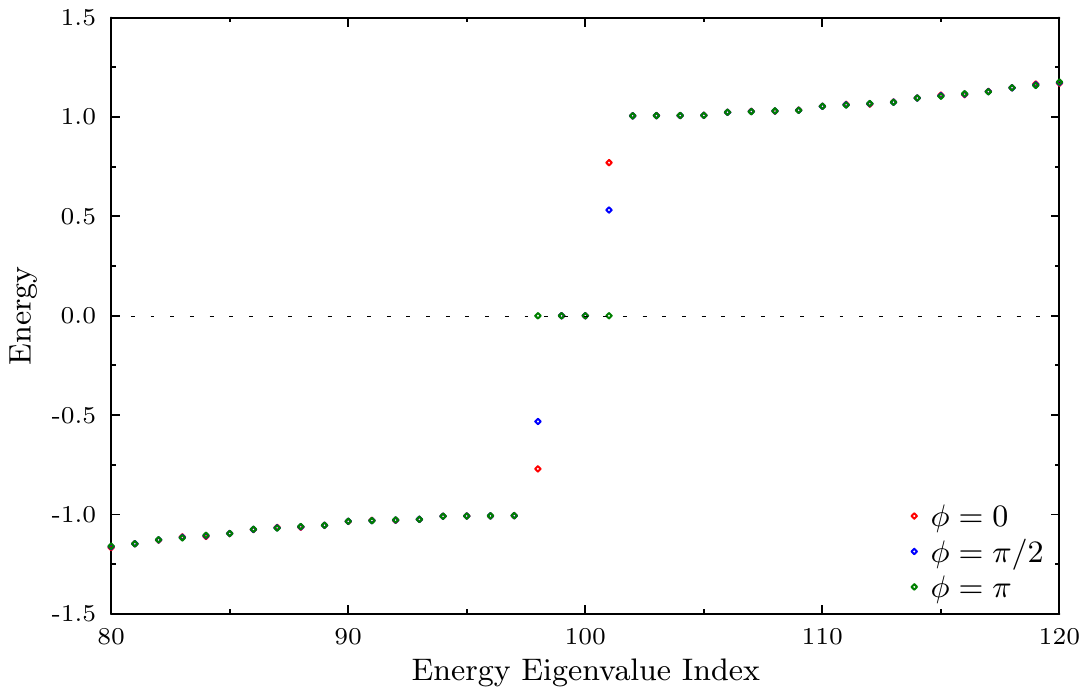}
\caption{(Color online) Top: The dependence of the lowest-energy eigenvalues on the SC phase difference between the two wires. Bottom: The eigenvalue spectrum for three different values of $\phi=0,\pi/2,\pi$. The system exhibits four zero-energy modes at $\phi=\pi$. The system parameters are $\lambda_1=1$ and $\lambda_2=1$.}
\label{fig:11}
\end{figure}

Let us first consider a Josephson junction between two wires with a
topological index $w^\alpha=1$. This type of junction has been extensively studied; here it is checked that the model yields results consistent with the known physics. The coupling constants are chosen to be $(\lambda_1^\alpha,\lambda_2^\alpha)=(1,1)$.
The numerical results for the dependence of the energy levels of this junction with the phase difference recover a zero-energy linear crossing at $\phi=\pi$ consistent with a $4 \pi$ periodicity in the anomalous Josephson effect (see Fig.~\ref{fig:11}).
The eigenvalue spectrum of the junction for $\phi=0,\pi/2,\pi$ is plotted in Fig.~\ref{fig:11}. Only for $\phi=\pi$ one recovers four zero-energy eigenvalues which correspond to four Majorana fermions: one at each extremity and two at the junction.

The $4\pi$ periodicity can be understood from a simple effective low-energy Hamiltonian.\cite{Kitaev}
The overlap between the wave functions of the two Majorana fermions at the extremities with the Majorana fermions at the interface is neglected. The simplest low-energy effective Hamiltonian can be written as\cite{Kitaev,Yakovenko04,Fu09,Lutchyn10,Alicea11,Ioselevich11,Badiane11,Jiang11}
\begin{equation}
H^{1-1}_{\rm eff}=i t_{12}\cos(\phi/2)\gamma_1\gamma_2,
\end{equation}
where $\gamma_1$ is a Majorana fermion at the right end of the first wire, $\gamma_2$ a Majorana fermion localized at the left end of the second wire and $t_{12}$ the effective tunneling amplitude (see Fig.~\ref{fig:junctions}). One can check that the cosine behavior reproduces well the low energy spectrum.
Introducing the fermion destruction operator $c=\frac{1}{\sqrt{2}}(\gamma_1+i\gamma_2)$, 
this is worth emphasizing that $i\gamma_1\gamma_2=c^\dag c-\frac{1}{2}$ is trivially a conserved quantity of $H^{1-1}_{\rm eff}$.

\subsection{The \texorpdfstring{$2-2$}{2-2} Josephson junction}
\label{subsec:22}
Let us  consider now a junction between two topological superconductors characterized by winding numbers
$w^\alpha=2$. 

\subsubsection{Analysis of the spectrum}
\begin{figure}[t]
\centering
\includegraphics[width=\columnwidth]{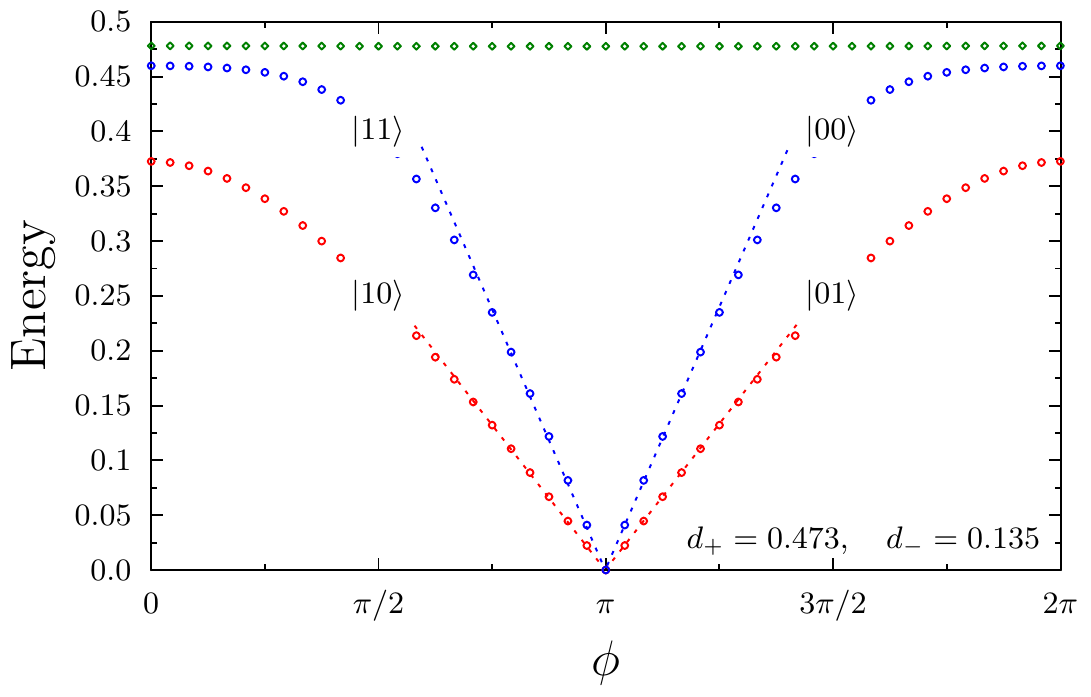}
\includegraphics[width=\columnwidth]{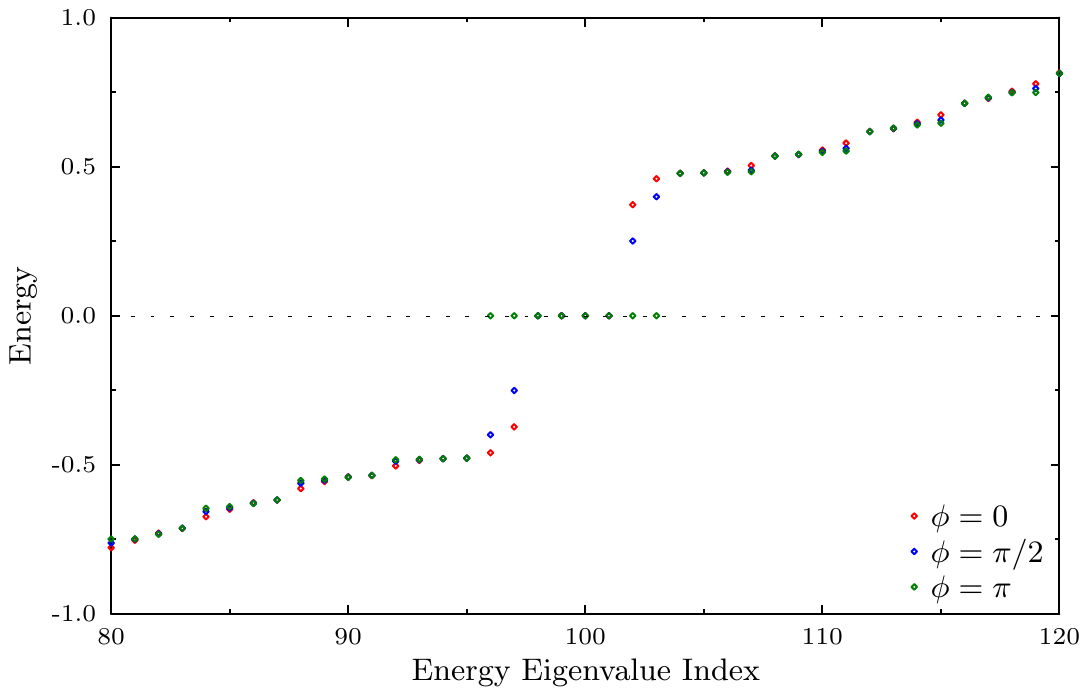}
\caption{(Color online.) Top: The dependence of the lowest-energy eigenvalues on the SC phase difference between the two wires. The energy branches are labeled using the occupation number representation for the two-fermion states formed from fusing four Majoranas across the junction. At low energy the two energy branches are linearly fitted: $E^{2-2}_+\pm E^{2-2}_-=(d_+\pm d_-)|\phi-\pi|/2$ and the value of coefficients $d_{\pm}$ is displayed.
Bottom: The eigenvalue spectrum for three different values of $\phi=0,\pi/2,\pi$. The system exhibits eight zero-energy modes at $\phi=\pi$. ($\lambda_1^{L,R}=1$, $\lambda_2^{L,R}=-1.5$).}
\label{fig:22}
\end{figure}

The eigenvalue spectrum for a $2-2$ junction is computed numerically. The result is shown in Fig.~\ref{fig:22}. The first important thing to note is that the anomalous $4\pi$ periodicity still holds. The only difference is that there are now four  Majorana fermions forming at the junction when the phase difference is $\phi=\pi$. At each extremity of the system there are also two Majorana fermions which subsist for any value of $\phi$, making the ground state eight-fold degenerate at $\phi=\pi$ (see Fig.~\ref{fig:22}).

A low-energy Hamiltonian capable to describe the numerical results presented in Fig.~\ref{fig:22} must involve four Majorana fermions. Let us denote by $\gamma_1,\gamma_2$
the Majorana fermions on the left side of the junction and by $\gamma_3,\gamma_4$ the Majorana fermions on the right side of the junction.
A general Hamiltonian involving the four Majoranas can be written as
\begin{eqnarray}\label{effH22}
H_{\rm eff}^{2-2}&=&i\gamma_1(t_{13}\gamma_3
+t_{14}\gamma_4)\cos\frac{\phi}{2}\notag\\
&&+i\gamma_2(t_{23}\gamma_3
+t_{24}\gamma_4)\cos\frac{\phi}{2}\\
&&+i( t_{34}\gamma_3\gamma_4+ t_{12}\gamma_1\gamma_2) f_{22}(\phi),\nonumber
\end{eqnarray}
where $f_{22}$ is an odd $2\pi$ periodic function of $\phi$.
The phase dependence of $H_{\rm eff}^{2-2}$ is fixed by enforcing a $2\pi$ periodicity together with the constraint that for $\phi=0$ there is no direct coupling between $\gamma_1$ and $\gamma_2$, nor between $\gamma_3$ and $\gamma_4$.\cite{Chakravarty}
The $\cos(\phi/2)$ is required by gauge invariance.
The effective Hamiltonian formally contains some direct tunneling terms between the Majorana fermions on the same side of the junctions ($t_{12}$ and $t_{34}$) which may be nonzero ($f_{22}(\phi)\ne 0$) when $\phi\ne n\pi$. However, one can argue that such terms must be less significant since they are the result of higher-order hopping processes between the two superconductors. Moreover, because they are odd functions of $\phi$, they are neglected in the following description of the low-energy spectrum at $\phi=\pi$.

Contrary to the $1-1$ junction, where there was an obvious quantity commuting with the Hamiltonian for all values of $\phi$, this quantity is not immediately evident in the $2-2$ junction. In the latter case, there are four Majorana fermions fusing to form two regular fermions that we denote by $c_\pm$. This can be seen from the spectral decomposition of the effective Hamiltonian in Eq.~(\ref{effH22}),
\begin{equation}\label{branches}
H_{\rm eff}^{2-2}=E_+^{2-2}(2c_+^\dag c_+^{\phantom\dag}-1)
+E^{2-2}_-(2c_-^\dag c_-^{\phantom\dag}-1),
\end{equation}
where
\begin{equation}
E_{\pm}^{2-2}=d_\pm\cos(\phi/2),
\end{equation}
with $d_\pm=\frac{1}{2\sqrt{2}}\sqrt{b\pm\sqrt{b^2-4a^2}}$ with $a=t_{14}t_{23}-t_{13}t_{24}$ and $b=t_{13}^2+t_{14}^2+t_{23}^2+t_{24}^2$. Note that even if the $f_{22}$ term in Eq.~(\ref{effH22}) was neglected, the form of the $c_{\pm}$ fermions as a function of the original Majorana fermions remains complicated. For example, under the reasonable assumption that the coupling strength between Majorana fermions situated at the same distance across the junction is identical, $t_{13}=t_{24}$, if follows that
\begin{eqnarray}
c_{\pm}&=&\frac{1}{\sqrt{2+2g^2_\pm}}[\gamma_4+i\gamma_1+g_{\pm}(\gamma_3+i\gamma_2)],\quad\text{with}\notag\\
g_\pm&=&\frac{1}{2t_{13}}\big[t_{23}-t_{14}\pm\sqrt{4t_{13}^2+(t_{14}-t_{23})^2}\big].
\end{eqnarray}
In this basis, the Hamiltonian in Eq. (\ref{branches}) commutes trivially with  the occupation numbers of the two fermions $n_\pm=c_\pm^\dag c^{\phantom\dag}_\pm$. Therefore, the occupation numbers of the two fermions are conserved and one can use the eigenvalues of the  $n_\pm$ operators to label the energy branches  of the two-particle states,
namely the  $|n_+n_-\rangle$ states as indicated in  Fig.~\ref{fig:22}. Such labeling of the four branches thus indicates that the level crossing is protected and hence guarantees the $4\pi$-periodicity of the dc Josephson effect for the 
$2-2$ junction. Concerning the tight-binding Hamiltonian,
the value of $E^{2-2}_\pm$ can be found numerically from a linear fit at low energy near $\phi=\pi$ in Fig.~\ref{fig:22}.

\subsubsection{Analysis of the Majorana polarization}
A useful tool to analyze the behavior of Majorana fermions is the Majorana polarization, a local topological order parameter introduced in Ref.~\onlinecite{Sticlet12}.
The Majorana polarization is proportional to the overlap of a wave function solution of the BdG equations with its particle-hole conjugate. This allows us to identify the presence of Majorana fermions at zero energy. Here, we define the $x$- and $y$-Majorana polarizations at zero energy and at site $i$ on a chain of length $N$ as 
\begin{eqnarray}
P^{(i)}_{M_x}(E)&=\frac{1}{2}\sum_{j=1}^{2N}\text{Re}\big[\Psi_j^{(i)}\tau_1\Psi_j^{(i)*}\big]e^{-E_j^2/w^2},\notag\\
P^{(i)}_{M_y}(E)&=\frac{1}{2}\sum_{j=1}^{2N}\text{Im}\big[\Psi_j^{(i)}\tau_1\Psi_j^{(i)*}\big]e^{-E_j^2/w^2},
\end{eqnarray}
where the sum is over all $(2N)$ energy eigenvalues. The eigenvectors $\Psi_j$ follow by solving numerically the mean-field BdG equation, $H_{BdG}\Psi_j=E_j\Psi_j$. The weight $w$ controls the spread of the energy eigenstates and is chosen in numerical simulation as $N^{-1}=10^{-2}$. 
The contribution of higher-energy states is thus exponentially suppressed. Subsequently the local Majorana polarization density is defined as the absolute value of the overlap, $P_M^{(i)2}=P_{M_x}^{(i)2}+P_{M_y}^{(i)2}$.
The existence of a Majorana fermion is recorded as a zero-energy Majorana polarization density of $0.5$ over a finite region of the wire. A Majorana fermion can have a $x$- and $y$-Majorana polarization. However, when the Hamiltonian is real, here at $\phi=0$ and $\phi=\pi$, the $y$-Majorana polarization $P_{M_y}$ vanishes. When the superconducting parameter acquires a phase, the polarization along the $y$ direction is generally nonzero.

\begin{figure}[t]
\centering
\includegraphics[width=\columnwidth]{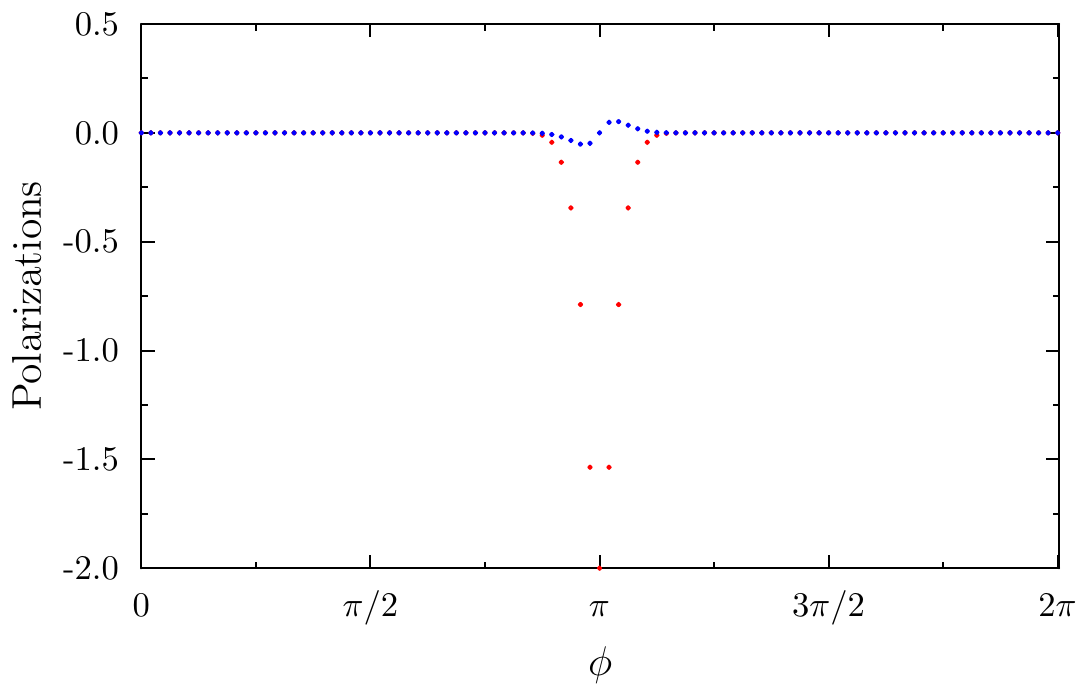}
\caption{(Color online) Zero-energy Majorana polarization ($x$-polarization in red and $y$- in blue) integrated over $40$ sites around the junction. Four Majorana states form at the junction for $\phi=\pi$ yielding a $x$-Majorana polarization of $2$, and compensating the Majorana polarization of the four end Majorana bound states (MBS).}
\label{fig:polz22mid}
\end{figure}

Here, similar to Ref.~\onlinecite{Chevalier12}, the end Majorana fermions respond to the variation of $\phi$ by rotating their Majorana polarization. This is due to the superconducting parameter becoming complex. However at $\phi=\pi$ the superconducting parameter becomes real again and the zero-energy end states have the same Majorana polarization. For this particular value of the phase, four new Majorana fermions form at the junction, two in each wire, their $x$-Majorana polarization compensating the Majorana polarization of the end modes (see Fig.~\ref{fig:polz22mid}).

\subsection{The \texorpdfstring{$1-2$}{1-2} Josephson junction}
\label{subsec:12}
In the following, we restrict our attention to the $1-2$ junction.
This is a particularly interesting problem since two distinct topological sectors are brought into contact via a Josephson junction. Here we assume that the left region is characterized by a winding number $w^L=1$ and has the couplings $\lambda^L_1=1$ and $\lambda^L_2=1$, while the right region, with $w^R=2$, has couplings $\lambda^R_1=1$, $\lambda^R_2=-1.5$.

\subsubsection{Analysis of the spectrum}
\begin{figure}[t]
\centering
\includegraphics[width=\columnwidth]{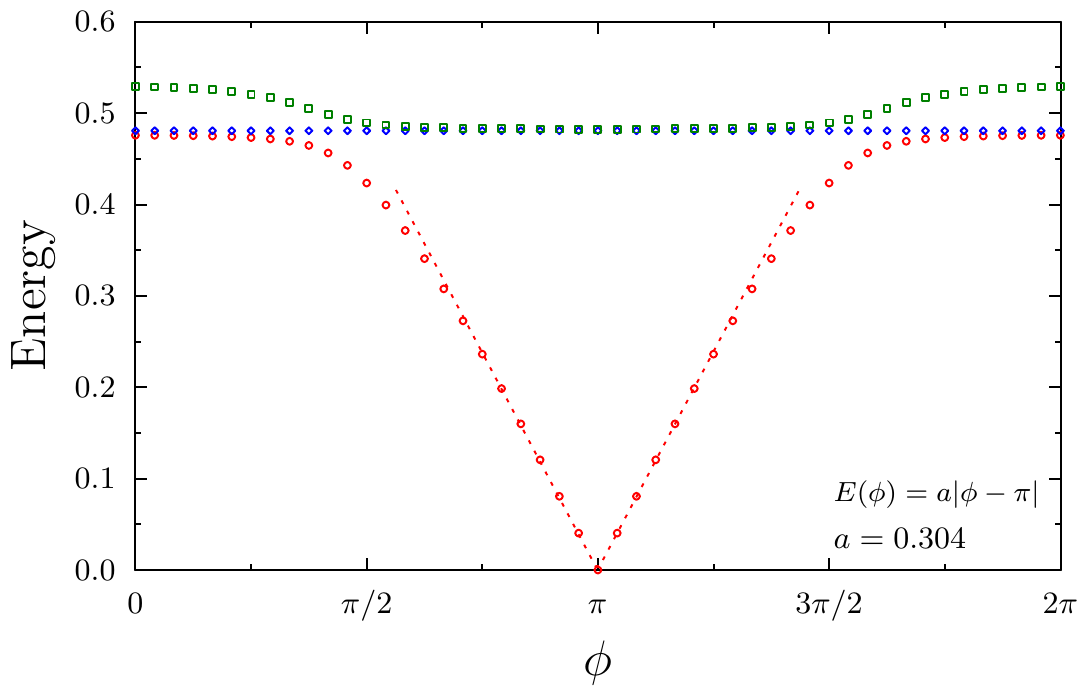} 
\includegraphics[width=\columnwidth]{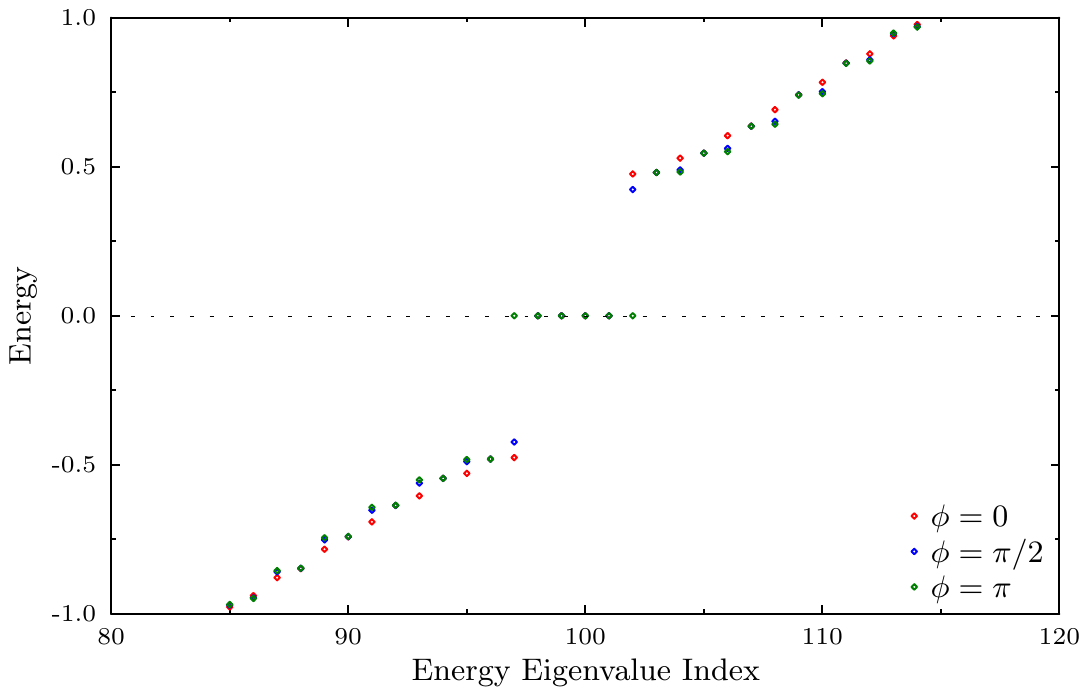} 
\caption{(Color online) Top: The dependence of the lowest-energy eigenvalues on the SC phase difference between the two wires. The dotted line indicate the fit of the lowest-energy numerical eigenvalue with the analytical form in $E_+$ from Eq.~(\ref{degenlift}). The fit parameter is $a=\sqrt{t^2_{12}+t^2_{13}}/4$. The zero-energy state independent of $\phi$ was not included in the figure.
Bottom: The eigenvalue spectrum for three different values of $\phi=0,\pi/2,\pi$. The system exhibits six zero-energy modes at $\phi=\pi$.}
\label{fig:12}
\end{figure}

It is expected that the physics of this junction will be dominated by three interacting Majorana fermions localized at the interface. One can show that for a winding number difference of 1 between the right and  left sides of the junction, one Majorana mode is bound at the interface for any choice of $\phi$.
Moreover, at $\phi=\pi$ the system is triply degenerate with three zero energy Majorana in the junction region (see Fig.~\ref{fig:12}).

Let $\gamma_1$ denote the Majorana fermion on the left side of the junction, and $\gamma_2,~\gamma_3$ the Majorana fermions on the right side. The gauge invariance of the Hamiltonian suggests that the phase-dependent couplings between $\gamma_1$ and $\gamma_{2,3}$ are proportional to $\cos(\phi/2)$ to compensate the sign change. 
A low-energy Hamiltonian describing the Majorana coupling in the $1-2$ system can thus be written as
\begin{equation}\label{effH}
H_{\rm eff}^{1-2}=i\gamma_1(t_{12}\gamma_2
+t_{13}\gamma_3)\cos\frac{\phi}{2}\\
+i t_{23}\gamma_2\gamma_3 f_{12}(\phi),
\end{equation}
where $f_{12}$ is an odd $2\pi$ periodic function of $\phi$.
At $\phi=\pi$ this Hamiltonian has three zero-energy eigenstates. For a phase difference of $\phi\neq\pi$, the effective Hamiltonian in (\ref{effH}) has one zero eigenvalue (required by the antisymmetry of the $3\times 3$ matrix)
and two non-zero eigenvalues. The constant zero-energy state, which is a Majorana edge state bound at the interface between two topologically nonequivalent regions, is the result of the difference of one unity between the topological indices of the two regions.

Similar to the $2-2$ junction, if the last term of Eq.~(\ref{effH}) is neglected, the form for the two non-zero eigenvalues becomes
\begin{equation}\label{degenlift}
E_{\pm}=\pm 2a\cos(\phi/2).
\end{equation}
The  hopping dependent parameter $a=\frac{1}{4}\sqrt{t^2_{12}+t^2_{13}}$ can be determined from a low-energy fit of the numerical dispersion presented in Fig.~\ref{fig:12}. Note that taking into account also the term $t_{23}$ improves the quality of the fit, especially in the vicinity of the superconducting gap. Note also that, while the fit is accurate up to energies close to the superconducting gap, the effective Hamiltonian $H_{\rm eff}^{1-2}$ is  only a low-energy effective Hamiltonian. Therefore,  one should not expect this Hamiltonian  to describe the full dependence of the energy eigenvalues with the SC phase difference.

\subsubsection{Majorana polarization analysis}
\begin{figure}[t]
\centering
\includegraphics[width=\columnwidth]{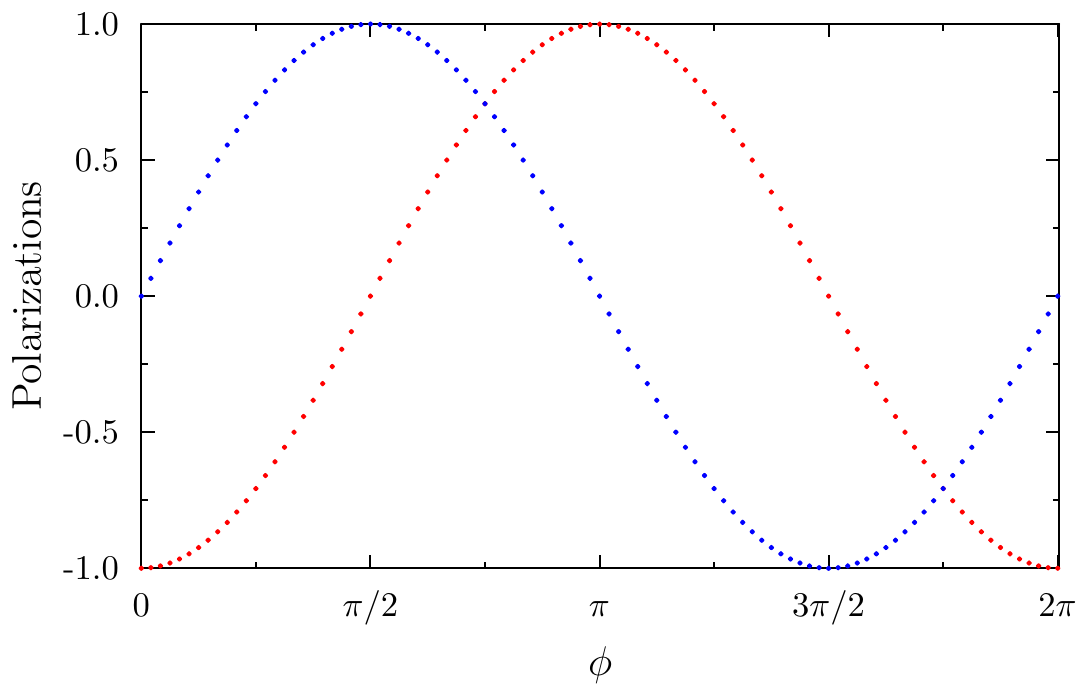}
\caption{(Color online.) The Majorana polarization of the right-end Majorana fermions integrated over the last $30$ sites. Note the oscillatory behavior of the $x$- (red) and $y$- (blue) Majorana polarization components with a total constant Majorana density of $2\times 0.5$, corresponding to two rotating Majorana modes.}
\label{fig:polzend}
\end{figure}

In the left-side wire, described by a winding number $w^L=1$, the superconducting parameter is chosen to be real and the single Majorana fermion at the left end is always fully $x$ polarized. For the right-hand-side wire, at the right end there are two Majorana fermions that respond to the twisting of the phase $\phi$ by gaining a $y$-polarization; The behavior of the $x$ and $y$ polarizations of these modes is presented in Fig.~\ref{fig:polzend}.

\begin{figure}[t]
\centering
\includegraphics[width=\columnwidth]{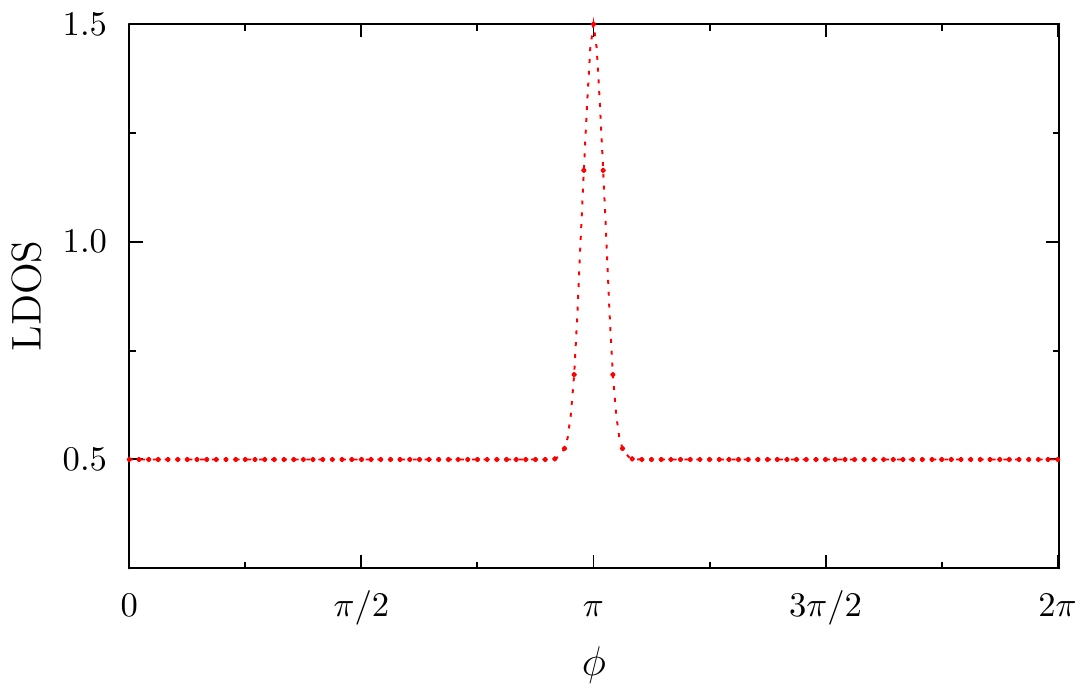}
\caption{Zero-energy density of states integrated over the $40$ sites around the junction point.}
\label{fig:ildos}
\end{figure}

\begin{figure}[t]
\centering
\includegraphics[width=\columnwidth]{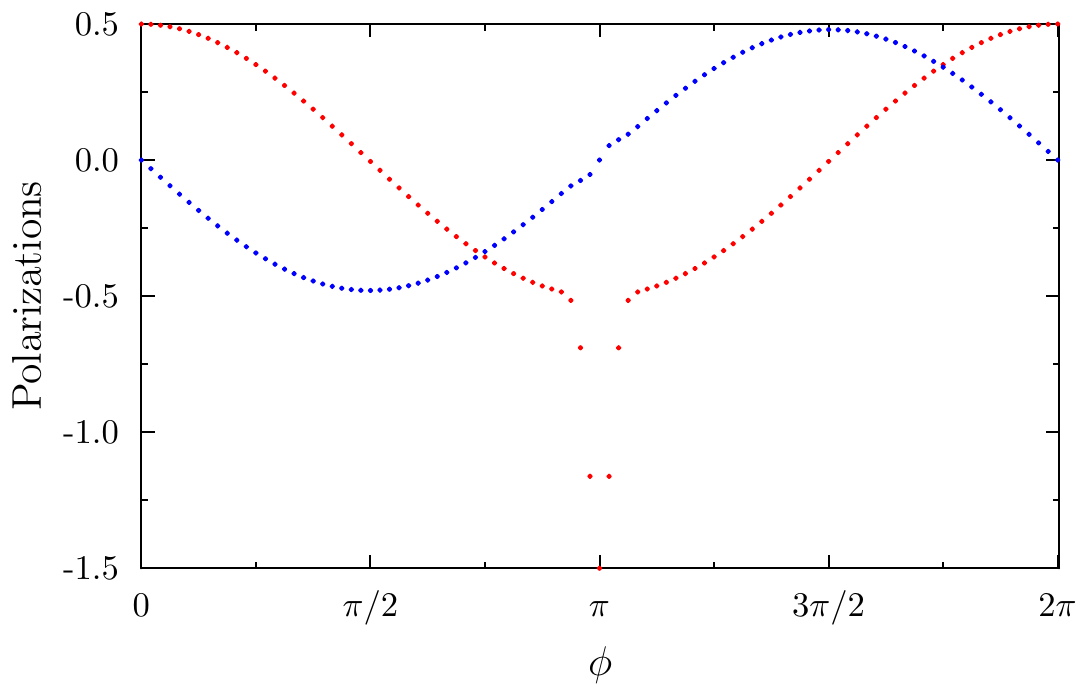}
\caption{(Color online.) Zero-energy Majorana polarization integrated over the $40$ sites around the junction. The $x$-polarization is depicted in red and the $y$-polarization in blue. Three Majorana fermions having the same polarization are supported at the junction for $\phi=\pi$.}
\label{fig:polz}
\end{figure}

Let us now analyze what happens at the junction between the two wires. At $\phi=\pi$, three Majorana fermions are expected. For any other value of $\phi$, there is always at least one Majorana fermion stuck at the interface between the two topologically nonequivalent regions. This can be seen by plotting  the zero-energy density of states, as well as the zero-energy Majorana polarization, integrated over the $40$ sites around the junction. In Fig.~\ref{fig:ildos} we plot the integrated zero-energy density of states. Note that for $\phi \ne \pi$ the density of states is constant and equal to $0.5$ corresponding to a single Majorana mode bound at the junction. The sharp jump between $0.5$ and $1.5$ at $\phi=\pi$ describes the contribution of two extra Majorana modes which reach zero energy at $\phi=\pi$ (these two extra zero-energy states appearing at $\phi=\pi$ can also be seen in the spectrum illustrated in Fig.~\ref{fig:12}).
This picture is confirmed by a plot of the zero-energy Majorana polarization, also integrated over the $40$ sites around the junction. The plot of the Majorana polarization is presented in Fig.~\ref{fig:polz}; the jump in the $x$-Majorana polarization at $\phi=\pi$ can be understood as coming from the two Majorana modes that develop at zero energy. Thus, at $\phi=\pi$ there are three Majorana fermions at the two extremities of the wire (one at the left end and two at the right end), fully $x$-polarized in the positive direction. These fermions are compensated by three Majorana fermions which form in the junction which are fully $x$-polarized in the opposite direction.  

\subsection{A general \texorpdfstring{$p-q$}{p-q} Josephson junction}
\label{subsec:pq}
In the following, we discuss the general case of a $p-q$ junction
between two wires, the left wire hosting $p$ Majorana fermions and the right one hosting $q$ Majorana
fermions. Without writing explicit tight-binding Hamiltonians, we treat low-energy effective Hamiltonians for  the $p-q$ junction:
\begin{equation}\label{effHpq}
H_{\rm eff}^{p-q}\simeq \sum\limits_{i=1}^p\sum\limits_{j=1}^q
it_{ij}\gamma_{L,i}\gamma_{R,j}\cos\frac{\phi}{2},
\end{equation}
where $\gamma_{L/R,k}$ labels the $k^{\rm th}$ Majorana fermion in the left/right lead. As before, the $\cos(\phi/2)$ is required by gauge invariance. The direct tunneling terms between Majorana fermions on the same side of the junction are neglected since they are the result of higher-order hopping processes between the two superconductors. Moreover, because these terms would involve odd functions of $\phi$, they can be neglected in the following description of the low-energy spectrum around $\phi=\pi$.

Two cases must be distinguished depending on the parity of $p+q$.
If $p+q$ is even, the spectral decomposition of the effective Hamiltonian in Eq.~\ref{effHpq}) can be written as 
\begin{equation}
H_{\rm eff}^{p-q}=\sum_{k=1}^{(p+q)/2} E^{p-q}_k (2c^\dag_kc_k-1) \cos\frac{\phi}{2},
\end{equation}
where the corresponding eigenvalues are complex functions of the
hopping variables $t_{ij}$, and the fermion operator $c_k$ is a linear combination of the Majorana operators $\gamma_{L,i}$ and $\gamma_{R,j}$.
The $p+q$ Majorana fermions can fuse to form at most $(p+q)/2$ complex fermions.
If $p+q$ is odd,  a similar analysis can be performed. Note however that in this case there is always at least one decoupled Majorana mode in the junction, and one can form at most $(p+q-1)/2$ complex fermions.

\section{Transport in SN junctions}
\label{sec:transp}
We have seen in the previous section that wires  supporting multiple Majorana fermions do support a $4\pi$ periodic Josephson effect. One may then wonder how to distinguish them
from wires supporting a single Majorana fermion. In this section, we show that this can be achieved using transport in SN junctions.

We focus in this section on an SN junction for which the SC is described by the Hamiltonian $H$ in Eq.~(\ref{eq:H}). For a junction between a wire with one Majorana end state and a normal metal, it has been predicted that the differential conductance exhibits a zero-bias peak of height $2e^2/h$.\cite{Law09,Flensberg10} A similar question for a junction between a topological superconducting wire characterized by a topological index $w>1$ and a normal metal has been recently addressed by Diez {\em et al.} in Ref.~\onlinecite{Diez12}. The authors  have shown that for such junctions the conductance $G$ can reach a value of $G=|w|\times 2e^2/h$. Here we check this prediction by considering a junction between a wire described by Eq.~(\ref{eq:H}) supporting four Majorana fermions, two at each of its extremities, and a normal wire.

The low-energy properties of such a system are determined by the coupling of the two end Majorana fermions with the normal metal. This coupling can be captured by a $2\times 2$ hybridization matrix $\bm\Gamma$.  
In order to compute the low-bias transport properties of this junction, 
one can directly use the $S$-matrix formalism developed by Flensberg,\cite{Flensberg10} and noting that the two wave functions for the Majorana fermions are orthogonal,\cite{Chakravarty} such that there is no inter-Majorana coupling term.

The resulting expression for the current can be written as \cite{Flensberg10}
\begin{equation}
I=\frac{e}{h}\int d\omega M(\omega)[f(-\omega+eV)-f(\omega-eV)],
\end{equation}
where $M(\omega)=\tr[\mb G^R(\omega)\bm\Gamma\mb G^A(\omega)\bm\Gamma(\omega)]$,
and
$\mb G^R(\omega)=2[\omega\mathbf 1+2i\bm\Gamma]^{-1}$
denotes the retarded Green's function.

The differential conductance becomes
\begin{equation}
\frac{dI}{dV}=-\frac{e^2}{h}\int d\omega M(\omega)
\bigg[\frac{df(-\omega+eV)}{d\omega}-\frac{df(\omega-eV)}{d\omega}
\bigg].
\end{equation}
which at $T=0$ reduces to
\begin{equation}
\frac{dI}{dV}=\frac{2e^2}{h}M(eV).
\end{equation}
Taking an explicit trace over the transmission matrix, we find 
\begin{equation}
\frac{dI}{dV}=\frac{8e^2}{h}
\frac{8\det(\bm\Gamma)^2+(eV)^2\tr(\bm\Gamma^2)}
{[(eV)^2-4\det(\bm\Gamma)]^2+[2eV\tr(\bm\Gamma)]^2}.
\end{equation}
which at zero bias becomes
\begin{equation}
\frac{dI}{dV}=\frac{4e^2}{h}.
\end{equation} 

The zero-bias value of the differential conductance is thus double that expected for a junction with a single Majorana fermion at the interface. This is consistent with each interface Majorana contributing a $2e^2/h$ to the total conductance. This line of reasoning can be straightforwardly extended to wires supporting an arbitrary large number of Majorana edge states.


\section{Wires with an inhomogeneous superconducting phase}
\label{sec:grad}
It was shown by Niu {\it et al.}\cite{Chakravarty} that, for a system with two Majorana modes at each end, the four zero modes are present provided the time-reversal and chiral symmetries are not broken, namely as long as the system is placed in the BDI class.\cite{Schnyder08,Schnyder09,Kitaev09}  Breaking time-reversal symmetry leads to the removal of the protection for the two additional Majorana fermions since the system now belongs to the D symmetry class. Then the system can return to the more typical state with at most one Majorana fermion at each end.

\begin{figure}[t]
\centering
\includegraphics[width=\columnwidth]{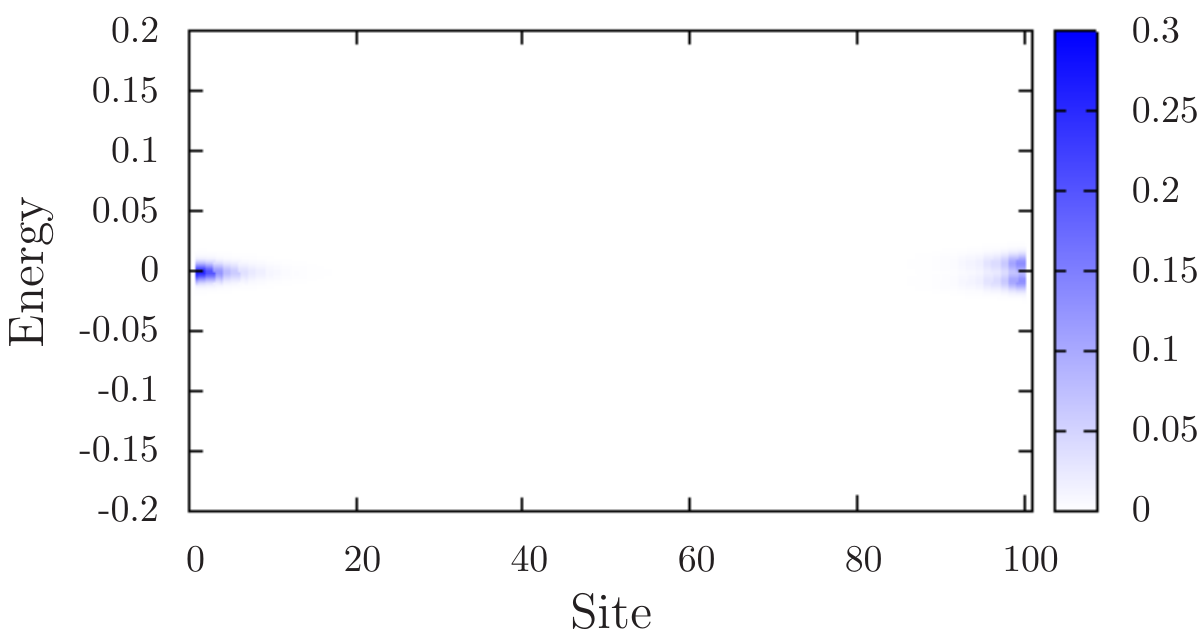}
\caption{Zero-energy local density of states. Addition of a small gradient $\nabla\phi=0.01$ per site leads to the removal of the protection of the zero modes and a hybridization of the right-end Majorana fermions.}
\label{fig:ldosgrad}
\end{figure}

Here the fragility of the topological phase is explored again by breaking the TRS through the action of a uniform superconducting phase gradient.
Let us take a wire characterized by the topological index $w=2$, with parameters $\lambda_1=1$ and $\lambda_2=-1.5$. A uniform phase gradient can be induced by the presence of supercurrents in the bulk of the superconductor. It was already shown that the phase gradient can be used to manipulate the creation or destruction of Majorana fermions.\cite{Romito} If the entire wire is subjected to the phase gradient, the BdG Hamiltonian gains a TRS breaking $h_1(k)$ odd component:
\begin{eqnarray}
h_1(k)&=&\lambda_1\sin(k)\sin(\nabla\phi/2)
+\lambda_2\sin(2k)\sin(\nabla\phi),\notag\\
h_2(k)&=&-\lambda_1\sin(k)\sin(\nabla\phi/2)
-\lambda_2\sin(2k)\cos(\nabla\phi), \\
h_3(k)&=&2-2\lambda_1\cos(k)\cos(\nabla\phi/2)
-2\lambda_2\cos(2k)\cos(\nabla\phi),\nonumber
\end{eqnarray} 
where the gradient $\nabla\phi$ is the change of the phase over one site. Note that a phase gradient also creates a nonvanishing $h_0$ component that multiplies the identity Pauli matrix.
Such a term tends to destroy the Cooper pairs. Let us assume a phase gradient, leading to an increase in the superconducting phase from left to right in a 100-site system. Because the Hamiltonian is now complex, the system is no longer characterized by a winding number $w$, but instead by the Kitaev ${\mathbb Z}_2$ invariant.\cite{Kitaev}
\begin{figure}[t]
\centering
\includegraphics[width=\columnwidth]{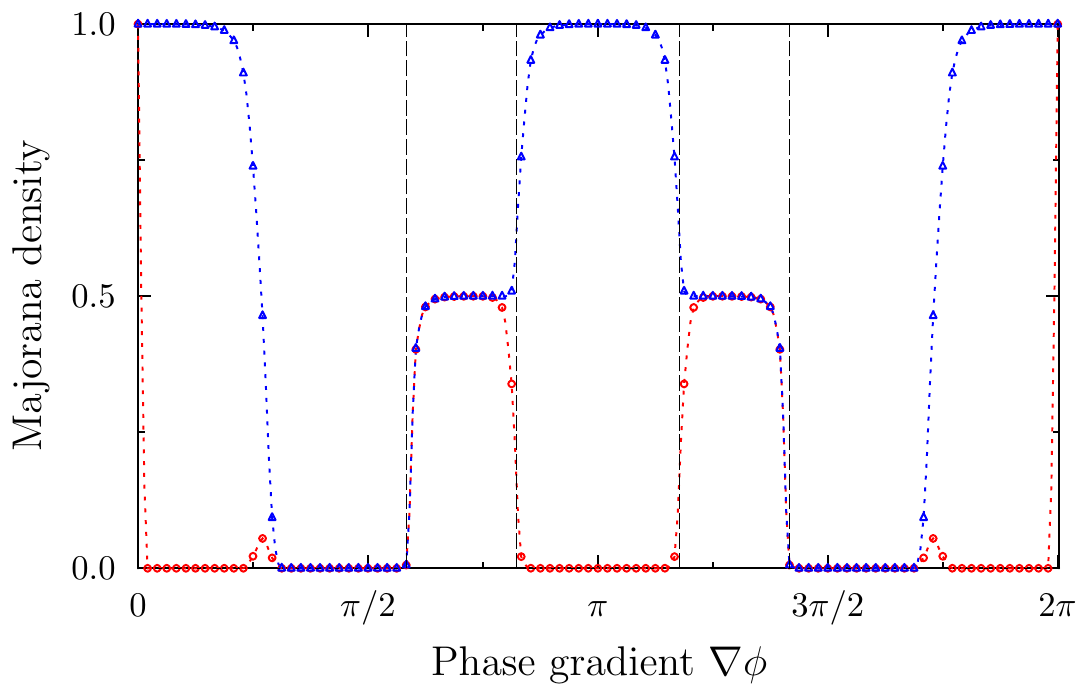}
\caption{(Color online) Zero-energy Majorana density integrated over 20 sites on the end left side (in blue) and on the right side (in red) of the wire. In the two regions predicted by by condition~(\ref{condition}) (which are delimited by the dashed lines for the chosen values of parameters $\lambda_1=1$, $\lambda_2=-1.5$) there is exactly one Majorana fermion at each side indicating a nontrivial $\mbb Z_2$ topological phase. However on the left side, there is a wide range of values of the gradient where there is still an unprotected Majorana doublet. 
}
\label{fig:mdens}
\end{figure}

If the entire wire experiences a uniform phase gradient, then it is in a topologically nontrivial $\mbb Z_2$ phase when the condition
\begin{equation}\label{condition}
|1-\lambda_2\cos(\nabla\phi)|<|\lambda_1\cos(\nabla\phi/2)|
\end{equation}
is satisfied.
Note that at vanishing phase gradient, one recovers the phase diagram in Fig.~\ref{fig:phdiag}, with the only change that $w=2$ and $w=0$ are both trivial $\mbb Z_2$ phases. Therefore the addition of a small gradient $\nabla\phi$ removes the protection of the Majorana doublets at the end of the wire. The accumulation of a SC phase near the left region of the wire leads to a split of the zero energy modes into regular fermionic states (see Fig.~\ref{fig:ldosgrad}). Note that for a strong enough phase gradient  obeying the condition in Eq. (\ref{condition}), the insulator makes a transition into a nontrivial $\mbb Z_2$ phase characterized by the presence of a single Majorana state at each end. This case is illustrated in Fig.~\ref{fig:mdens} where we plot the zero-energy Majorana density integrated over the last 20 sites of the right and left side of the wire as a function of the gradient. The phase gradient generally destroys the phase with multiple Majorana fermions since it renders complex the Hamiltonian which describes the wire. Note however that when the gradient is close to $0$ or $\pi$ there is an unusual persistence of a doublet of Majorana fermions on the left side of the wire. This might be explained by observing that the superconducting phase starts at zero on the left side of the wire and the superconducting parameters have only a small complex component for $\nabla\phi\simeq 0\text{ or }\pi$. Then the Majorana doublet could be understood as a remnant of the $\mbb Z$ phase for a small chiral symmetry breaking.    
Nevertheless, the general conclusion remains valid: the phase gradient can serve as an external knob to switch from a $\mbb Z$ topological superconductor to either a trivial or non-trivial $\mbb Z_2$ insulator.
\section{Conclusions}
\label{sec:conc}
In this paper, we have studied in detail the Josephson effect between narrow semiconducting wires  described by  band Hamiltonians in the BDI symmetry class  which
can support multiple Majorana end modes. Such situation may be encountered in relatively narrow semiconducting wires with a strong spin-orbit coupling, in the presence of a Zeeman splitting and in the proximity of an $s$-wave superconductor.

Following  Niu {\it et al.}, \cite{Chakravarty} we mainly focused on two-band Hamiltonians resulting from  nearest- and next-nearest-neighbor hopping integrals in a 
tight-binding description of such narrow wires.
Based on exact diagonalization of a tight-binding model hosting multiple Majorana fermions together with a 
low-energy description of the junction, we have shown that the $4\pi$ anomalous periodicity of the dc Josephson effect is maintained. Even though the present work considered junctions between topological wires supporting one or two Majorana fermions at their extremities, the low-energy description can be straightforwardly extended to junctions between wires supporting an arbitrary number of Majorana fermions.

Finally, we have shown that a uniform superconducting phase gradient is able to lift the protection for the states with multiple Majorana fermions and push the system into a nontrivial $\mbb Z_2$ state characterized by the presence of a unique Majorana state at each end.

The authors wish to thank J.\ Alicea and R.\ Aguado for stimulating discussions. The work of C.B. was supported by ERC Starting Independent Researcher Grant No. NANO-GRAPHENE 256965.

\bibliographystyle{apsrev4-1}
\bibliography{bibl}

\begin{thebibliography}{40}%
\makeatletter
\providecommand \@ifxundefined [1]{%
 \@ifx{#1\undefined}
}%
\providecommand \@ifnum [1]{%
 \ifnum #1\expandafter \@firstoftwo
 \else \expandafter \@secondoftwo
 \fi
}%
\providecommand \@ifx [1]{%
 \ifx #1\expandafter \@firstoftwo
 \else \expandafter \@secondoftwo
 \fi
}%
\providecommand \natexlab [1]{#1}%
\providecommand \enquote  [1]{``#1''}%
\providecommand \bibnamefont  [1]{#1}%
\providecommand \bibfnamefont [1]{#1}%
\providecommand \citenamefont [1]{#1}%
\providecommand \href@noop [0]{\@secondoftwo}%
\providecommand \href [0]{\begingroup \@sanitize@url \@href}%
\providecommand \@href[1]{\@@startlink{#1}\@@href}%
\providecommand \@@href[1]{\endgroup#1\@@endlink}%
\providecommand \@sanitize@url [0]{\catcode `\\12\catcode `\$12\catcode
  `\&12\catcode `\#12\catcode `\^12\catcode `\_12\catcode `\%12\relax}%
\providecommand \@@startlink[1]{}%
\providecommand \@@endlink[0]{}%
\providecommand \url  [0]{\begingroup\@sanitize@url \@url }%
\providecommand \@url [1]{\endgroup\@href {#1}{\urlprefix }}%
\providecommand \urlprefix  [0]{URL }%
\providecommand \Eprint [0]{\href }%
\providecommand \doibase [0]{http://dx.doi.org/}%
\providecommand \selectlanguage [0]{\@gobble}%
\providecommand \bibinfo  [0]{\@secondoftwo}%
\providecommand \bibfield  [0]{\@secondoftwo}%
\providecommand \translation [1]{[#1]}%
\providecommand \BibitemOpen [0]{}%
\providecommand \bibitemStop [0]{}%
\providecommand \bibitemNoStop [0]{.\EOS\space}%
\providecommand \EOS [0]{\spacefactor3000\relax}%
\providecommand \BibitemShut  [1]{\csname bibitem#1\endcsname}%
\let\auto@bib@innerbib\@empty
\bibitem [{\citenamefont {Alicea}(2012)}]{Alicea12}%
  \BibitemOpen
  \bibfield  {author} {\bibinfo {author} {\bibfnamefont {J.}~\bibnamefont
  {Alicea}},\ }\href {\doibase 10.1088/0034-4885/75/7/076501} {\bibfield
  {journal} {\bibinfo  {journal} {Rep. Prog. Phys.}\ }\textbf {\bibinfo
  {volume} {75}},\ \bibinfo {pages} {076501} (\bibinfo {year}
  {2012})}\BibitemShut {NoStop}%
\bibitem [{\citenamefont {{Beenakker}}(2011)}]{Beenakker12}%
  \BibitemOpen
  \bibfield  {author} {\bibinfo {author} {\bibfnamefont {C.~W.~J.}\
  \bibnamefont {{Beenakker}}},\ }\href@noop {} {\  (\bibinfo {year} {2011})},\
  \Eprint {http://arxiv.org/abs/1112.1950} {arXiv:1112.1950
  [cond-mat.mes-hall]} \BibitemShut {NoStop}%
\bibitem [{\citenamefont {Nayak}\ \emph {et~al.}(2008)\citenamefont {Nayak},
  \citenamefont {Simon}, \citenamefont {Stern}, \citenamefont {Freedman},\ and\
  \citenamefont {Das~Sarma}}]{Nayak08}%
  \BibitemOpen
  \bibfield  {author} {\bibinfo {author} {\bibfnamefont {C.}~\bibnamefont
  {Nayak}}, \bibinfo {author} {\bibfnamefont {S.~H.}\ \bibnamefont {Simon}},
  \bibinfo {author} {\bibfnamefont {A.}~\bibnamefont {Stern}}, \bibinfo
  {author} {\bibfnamefont {M.}~\bibnamefont {Freedman}}, \ and\ \bibinfo
  {author} {\bibfnamefont {S.}~\bibnamefont {Das~Sarma}},\ }\href {\doibase
  10.1103/RevModPhys.80.1083} {\bibfield  {journal} {\bibinfo  {journal} {Rev.
  Mod. Phys.}\ }\textbf {\bibinfo {volume} {80}},\ \bibinfo {pages} {1083}
  (\bibinfo {year} {2008})}\BibitemShut {NoStop}%
\bibitem [{\citenamefont {Mourik}\ \emph {et~al.}(2012)\citenamefont {Mourik},
  \citenamefont {Zuo}, \citenamefont {Frolov}, \citenamefont {Plissard},
  \citenamefont {Bakkers},\ and\ \citenamefont {Kouwenhoven}}]{Mourik12}%
  \BibitemOpen
  \bibfield  {author} {\bibinfo {author} {\bibfnamefont {V.}~\bibnamefont
  {Mourik}}, \bibinfo {author} {\bibfnamefont {K.}~\bibnamefont {Zuo}},
  \bibinfo {author} {\bibfnamefont {S.~M.}\ \bibnamefont {Frolov}}, \bibinfo
  {author} {\bibfnamefont {S.~R.}\ \bibnamefont {Plissard}}, \bibinfo {author}
  {\bibfnamefont {E.~P. A.~M.}\ \bibnamefont {Bakkers}}, \ and\ \bibinfo
  {author} {\bibfnamefont {L.~P.}\ \bibnamefont {Kouwenhoven}},\ }\href
  {\doibase 10.1126/science.1222360} {\bibfield  {journal} {\bibinfo  {journal}
  {Science}\ }\textbf {\bibinfo {volume} {336}},\ \bibinfo {pages} {1003}
  (\bibinfo {year} {2012})}\BibitemShut {NoStop}%
\bibitem [{\citenamefont {{Deng}}\ \emph {et~al.}(2012)\citenamefont {{Deng}},
  \citenamefont {{Yu}}, \citenamefont {{Huang}}, \citenamefont {{Larsson}},
  \citenamefont {{Caroff}},\ and\ \citenamefont {{Xu}}}]{Xu12}%
  \BibitemOpen
  \bibfield  {author} {\bibinfo {author} {\bibfnamefont {M.~T.}\ \bibnamefont
  {{Deng}}}, \bibinfo {author} {\bibfnamefont {C.~L.}\ \bibnamefont {{Yu}}},
  \bibinfo {author} {\bibfnamefont {G.~Y.}\ \bibnamefont {{Huang}}}, \bibinfo
  {author} {\bibfnamefont {M.}~\bibnamefont {{Larsson}}}, \bibinfo {author}
  {\bibfnamefont {P.}~\bibnamefont {{Caroff}}}, \ and\ \bibinfo {author}
  {\bibfnamefont {H.~Q.}\ \bibnamefont {{Xu}}},\ }\href@noop {} {\bibfield
  {journal} {\bibinfo  {journal} {Nano Lett.}\ }\textbf {\bibinfo {volume}
  {12}},\ \bibinfo {pages} {6414} (\bibinfo {year} {2012})}\BibitemShut
  {NoStop}%
\bibitem [{\citenamefont {{Das}}\ \emph {et~al.}(2012)\citenamefont {{Das}},
  \citenamefont {{Ronen}}, \citenamefont {{Most}}, \citenamefont {{Oreg}},
  \citenamefont {{Heiblum}},\ and\ \citenamefont {{Shtrikman}}}]{Das12}%
  \BibitemOpen
  \bibfield  {author} {\bibinfo {author} {\bibfnamefont {A.}~\bibnamefont
  {{Das}}}, \bibinfo {author} {\bibfnamefont {Y.}~\bibnamefont {{Ronen}}},
  \bibinfo {author} {\bibfnamefont {Y.}~\bibnamefont {{Most}}}, \bibinfo
  {author} {\bibfnamefont {Y.}~\bibnamefont {{Oreg}}}, \bibinfo {author}
  {\bibfnamefont {M.}~\bibnamefont {{Heiblum}}}, \ and\ \bibinfo {author}
  {\bibfnamefont {H.}~\bibnamefont {{Shtrikman}}},\ }\href@noop {} {\bibfield
  {journal} {\bibinfo  {journal} {Nature Physics}\ }\textbf {\bibinfo {volume}
  {8}},\ \bibinfo {pages} {887} (\bibinfo {year} {2012})}\BibitemShut {NoStop}%
\bibitem [{\citenamefont {Bagrets}\ and\ \citenamefont
  {Altland}(2012)}]{Bagrets12}%
  \BibitemOpen
  \bibfield  {author} {\bibinfo {author} {\bibfnamefont {D.}~\bibnamefont
  {Bagrets}}\ and\ \bibinfo {author} {\bibfnamefont {A.}~\bibnamefont
  {Altland}},\ }\href {\doibase 10.1103/PhysRevLett.109.227005} {\bibfield
  {journal} {\bibinfo  {journal} {Phys. Rev. Lett.}\ }\textbf {\bibinfo
  {volume} {109}},\ \bibinfo {pages} {227005} (\bibinfo {year}
  {2012})}\BibitemShut {NoStop}%
\bibitem [{\citenamefont {Liu}\ \emph {et~al.}(2012)\citenamefont {Liu},
  \citenamefont {Potter}, \citenamefont {Law},\ and\ \citenamefont
  {Lee}}]{Liu12}%
  \BibitemOpen
  \bibfield  {author} {\bibinfo {author} {\bibfnamefont {J.}~\bibnamefont
  {Liu}}, \bibinfo {author} {\bibfnamefont {A.~C.}\ \bibnamefont {Potter}},
  \bibinfo {author} {\bibfnamefont {K.~T.}\ \bibnamefont {Law}}, \ and\
  \bibinfo {author} {\bibfnamefont {P.~A.}\ \bibnamefont {Lee}},\ }\href
  {\doibase 10.1103/PhysRevLett.109.267002} {\bibfield  {journal} {\bibinfo
  {journal} {Phys. Rev. Lett.}\ }\textbf {\bibinfo {volume} {109}},\ \bibinfo
  {pages} {267002} (\bibinfo {year} {2012})}\BibitemShut {NoStop}%
\bibitem [{\citenamefont {Pientka}\ \emph {et~al.}(2012)\citenamefont
  {Pientka}, \citenamefont {Kells}, \citenamefont {Romito}, \citenamefont
  {Brouwer},\ and\ \citenamefont {von Oppen}}]{Pientka12}%
  \BibitemOpen
  \bibfield  {author} {\bibinfo {author} {\bibfnamefont {F.}~\bibnamefont
  {Pientka}}, \bibinfo {author} {\bibfnamefont {G.}~\bibnamefont {Kells}},
  \bibinfo {author} {\bibfnamefont {A.}~\bibnamefont {Romito}}, \bibinfo
  {author} {\bibfnamefont {P.~W.}\ \bibnamefont {Brouwer}}, \ and\ \bibinfo
  {author} {\bibfnamefont {F.}~\bibnamefont {von Oppen}},\ }\href {\doibase
  10.1103/PhysRevLett.109.227006} {\bibfield  {journal} {\bibinfo  {journal}
  {Phys. Rev. Lett.}\ }\textbf {\bibinfo {volume} {109}},\ \bibinfo {pages}
  {227006} (\bibinfo {year} {2012})}\BibitemShut {NoStop}%
\bibitem [{\citenamefont {Rainis}\ \emph {et~al.}(2013)\citenamefont {Rainis},
  \citenamefont {Trifunovic}, \citenamefont {Klinovaja},\ and\ \citenamefont
  {Loss}}]{Rainis12}%
  \BibitemOpen
  \bibfield  {author} {\bibinfo {author} {\bibfnamefont {D.}~\bibnamefont
  {Rainis}}, \bibinfo {author} {\bibfnamefont {L.}~\bibnamefont {Trifunovic}},
  \bibinfo {author} {\bibfnamefont {J.}~\bibnamefont {Klinovaja}}, \ and\
  \bibinfo {author} {\bibfnamefont {D.}~\bibnamefont {Loss}},\ }\href {\doibase
  10.1103/PhysRevB.87.024515} {\bibfield  {journal} {\bibinfo  {journal} {Phys.
  Rev. B}\ }\textbf {\bibinfo {volume} {87}},\ \bibinfo {pages} {024515}
  (\bibinfo {year} {2013})}\BibitemShut {NoStop}%
\bibitem [{\citenamefont {{Kitaev}}(2001)}]{Kitaev}%
  \BibitemOpen
  \bibfield  {author} {\bibinfo {author} {\bibfnamefont {A.~Y.}\ \bibnamefont
  {{Kitaev}}},\ }\href {\doibase 10.1070/1063-7869/44/10S/S29} {\bibfield
  {journal} {\bibinfo  {journal} {Physics Uspekhi}\ }\textbf {\bibinfo {volume}
  {44}},\ \bibinfo {pages} {131} (\bibinfo {year} {2001})},\ \Eprint
  {http://arxiv.org/abs/cond-mat/0010440} {arXiv:cond-mat/0010440} \BibitemShut
  {NoStop}%
\bibitem [{\citenamefont {{Kwon}}\ \emph {et~al.}(2003)\citenamefont {{Kwon}},
  \citenamefont {{Sengupta}},\ and\ \citenamefont {{Yakovenko}}}]{Kwon03}%
  \BibitemOpen
  \bibfield  {author} {\bibinfo {author} {\bibfnamefont {H.-J.}\ \bibnamefont
  {{Kwon}}}, \bibinfo {author} {\bibfnamefont {K.}~\bibnamefont {{Sengupta}}},
  \ and\ \bibinfo {author} {\bibfnamefont {V.~M.}\ \bibnamefont
  {{Yakovenko}}},\ }\href@noop {} {\bibfield  {journal} {\bibinfo  {journal}
  {European Physical Journal B}\ }\textbf {\bibinfo {volume} {37}},\ \bibinfo
  {pages} {349} (\bibinfo {year} {2003})}\BibitemShut {NoStop}%
\bibitem [{\citenamefont {{Kwon}}\ \emph {et~al.}(2004)\citenamefont {{Kwon}},
  \citenamefont {{Yakovenko}},\ and\ \citenamefont {{Sengupta}}}]{Yakovenko04}%
  \BibitemOpen
  \bibfield  {author} {\bibinfo {author} {\bibfnamefont {H.-J.}\ \bibnamefont
  {{Kwon}}}, \bibinfo {author} {\bibfnamefont {V.~M.}\ \bibnamefont
  {{Yakovenko}}}, \ and\ \bibinfo {author} {\bibfnamefont {K.}~\bibnamefont
  {{Sengupta}}},\ }\href {\doibase 10.1063/1.1789931} {\bibfield  {journal}
  {\bibinfo  {journal} {Low Temperature Physics}\ }\textbf {\bibinfo {volume}
  {30}},\ \bibinfo {pages} {613} (\bibinfo {year} {2004})},\ \Eprint
  {http://arxiv.org/abs/arXiv:0401313} {arXiv:0401313 [cond-mat]} \BibitemShut
  {NoStop}%
\bibitem [{\citenamefont {Fu}\ and\ \citenamefont {Kane}(2009)}]{Fu09}%
  \BibitemOpen
  \bibfield  {author} {\bibinfo {author} {\bibfnamefont {L.}~\bibnamefont
  {Fu}}\ and\ \bibinfo {author} {\bibfnamefont {C.~L.}\ \bibnamefont {Kane}},\
  }\href {\doibase 10.1103/PhysRevB.79.161408} {\bibfield  {journal} {\bibinfo
  {journal} {Phys. Rev. B}\ }\textbf {\bibinfo {volume} {79}},\ \bibinfo
  {pages} {161408} (\bibinfo {year} {2009})}\BibitemShut {NoStop}%
\bibitem [{\citenamefont {Badiane}\ \emph {et~al.}(2011)\citenamefont
  {Badiane}, \citenamefont {Houzet},\ and\ \citenamefont {Meyer}}]{Badiane11}%
  \BibitemOpen
  \bibfield  {author} {\bibinfo {author} {\bibfnamefont {D.~M.}\ \bibnamefont
  {Badiane}}, \bibinfo {author} {\bibfnamefont {M.}~\bibnamefont {Houzet}}, \
  and\ \bibinfo {author} {\bibfnamefont {J.~S.}\ \bibnamefont {Meyer}},\ }\href
  {\doibase 10.1103/PhysRevLett.107.177002} {\bibfield  {journal} {\bibinfo
  {journal} {Phys. Rev. Lett.}\ }\textbf {\bibinfo {volume} {107}},\ \bibinfo
  {pages} {177002} (\bibinfo {year} {2011})}\BibitemShut {NoStop}%
\bibitem [{\citenamefont {Ioselevich}\ and\ \citenamefont
  {Feigel'man}(2011)}]{Ioselevich11}%
  \BibitemOpen
  \bibfield  {author} {\bibinfo {author} {\bibfnamefont {P.~A.}\ \bibnamefont
  {Ioselevich}}\ and\ \bibinfo {author} {\bibfnamefont {M.~V.}\ \bibnamefont
  {Feigel'man}},\ }\href {\doibase 10.1103/PhysRevLett.106.077003} {\bibfield
  {journal} {\bibinfo  {journal} {Phys. Rev. Lett.}\ }\textbf {\bibinfo
  {volume} {106}},\ \bibinfo {pages} {077003} (\bibinfo {year}
  {2011})}\BibitemShut {NoStop}%
\bibitem [{\citenamefont {Alicea}\ \emph {et~al.}(2011)\citenamefont {Alicea},
  \citenamefont {Oreg}, \citenamefont {Refael}, \citenamefont {von Oppen},\
  and\ \citenamefont {Fisher}}]{Alicea11}%
  \BibitemOpen
  \bibfield  {author} {\bibinfo {author} {\bibfnamefont {J.}~\bibnamefont
  {Alicea}}, \bibinfo {author} {\bibfnamefont {Y.}~\bibnamefont {Oreg}},
  \bibinfo {author} {\bibfnamefont {G.}~\bibnamefont {Refael}}, \bibinfo
  {author} {\bibfnamefont {F.}~\bibnamefont {von Oppen}}, \ and\ \bibinfo
  {author} {\bibfnamefont {M.~P.~A.}\ \bibnamefont {Fisher}},\ }\href {\doibase
  10.1038/nphys1915} {\bibfield  {journal} {\bibinfo  {journal} {Nature
  Physics}\ }\textbf {\bibinfo {volume} {7}},\ \bibinfo {pages} {412} (\bibinfo
  {year} {2011})}\BibitemShut {NoStop}%
\bibitem [{\citenamefont {Jiang}\ \emph {et~al.}(2011)\citenamefont {Jiang},
  \citenamefont {Pekker}, \citenamefont {Alicea}, \citenamefont {Refael},
  \citenamefont {Oreg},\ and\ \citenamefont {von Oppen}}]{Jiang11}%
  \BibitemOpen
  \bibfield  {author} {\bibinfo {author} {\bibfnamefont {L.}~\bibnamefont
  {Jiang}}, \bibinfo {author} {\bibfnamefont {D.}~\bibnamefont {Pekker}},
  \bibinfo {author} {\bibfnamefont {J.}~\bibnamefont {Alicea}}, \bibinfo
  {author} {\bibfnamefont {G.}~\bibnamefont {Refael}}, \bibinfo {author}
  {\bibfnamefont {Y.}~\bibnamefont {Oreg}}, \ and\ \bibinfo {author}
  {\bibfnamefont {F.}~\bibnamefont {von Oppen}},\ }\href {\doibase
  10.1103/PhysRevLett.107.236401} {\bibfield  {journal} {\bibinfo  {journal}
  {Phys. Rev. Lett.}\ }\textbf {\bibinfo {volume} {107}},\ \bibinfo {pages}
  {236401} (\bibinfo {year} {2011})}\BibitemShut {NoStop}%
\bibitem [{\citenamefont {San-Jose}\ \emph {et~al.}(2012)\citenamefont
  {San-Jose}, \citenamefont {Prada},\ and\ \citenamefont
  {Aguado}}]{San-Jose12}%
  \BibitemOpen
  \bibfield  {author} {\bibinfo {author} {\bibfnamefont {P.}~\bibnamefont
  {San-Jose}}, \bibinfo {author} {\bibfnamefont {E.}~\bibnamefont {Prada}}, \
  and\ \bibinfo {author} {\bibfnamefont {R.}~\bibnamefont {Aguado}},\ }\href
  {\doibase 10.1103/PhysRevLett.108.257001} {\bibfield  {journal} {\bibinfo
  {journal} {Phys. Rev. Lett.}\ }\textbf {\bibinfo {volume} {108}},\ \bibinfo
  {pages} {257001} (\bibinfo {year} {2012})}\BibitemShut {NoStop}%
\bibitem [{\citenamefont {Pikulin}\ and\ \citenamefont
  {Nazarov}(2012)}]{Pikulin12}%
  \BibitemOpen
  \bibfield  {author} {\bibinfo {author} {\bibfnamefont {D.~I.}\ \bibnamefont
  {Pikulin}}\ and\ \bibinfo {author} {\bibfnamefont {Y.~V.}\ \bibnamefont
  {Nazarov}},\ }\href {\doibase 10.1103/PhysRevB.86.140504} {\bibfield
  {journal} {\bibinfo  {journal} {Phys. Rev. B}\ }\textbf {\bibinfo {volume}
  {86}},\ \bibinfo {pages} {140504} (\bibinfo {year} {2012})}\BibitemShut
  {NoStop}%
\bibitem [{\citenamefont {Dominguez}\ \emph {et~al.}(2012)\citenamefont
  {Dominguez}, \citenamefont {Hassler},\ and\ \citenamefont
  {Platero}}]{Platero12}%
  \BibitemOpen
  \bibfield  {author} {\bibinfo {author} {\bibfnamefont {F.}~\bibnamefont
  {Dominguez}}, \bibinfo {author} {\bibfnamefont {F.}~\bibnamefont {Hassler}},
  \ and\ \bibinfo {author} {\bibfnamefont {G.}~\bibnamefont {Platero}},\ }\href
  {\doibase 10.1103/PhysRevB.86.140503} {\bibfield  {journal} {\bibinfo
  {journal} {Phys. Rev. B}\ }\textbf {\bibinfo {volume} {86}},\ \bibinfo
  {pages} {140503} (\bibinfo {year} {2012})}\BibitemShut {NoStop}%
\bibitem [{\citenamefont {Law}\ and\ \citenamefont {Lee}(2011)}]{Law11}%
  \BibitemOpen
  \bibfield  {author} {\bibinfo {author} {\bibfnamefont {K.~T.}\ \bibnamefont
  {Law}}\ and\ \bibinfo {author} {\bibfnamefont {P.~A.}\ \bibnamefont {Lee}},\
  }\href {\doibase 10.1103/PhysRevB.84.081304} {\bibfield  {journal} {\bibinfo
  {journal} {Phys. Rev. B}\ }\textbf {\bibinfo {volume} {84}},\ \bibinfo
  {pages} {081304} (\bibinfo {year} {2011})}\BibitemShut {NoStop}%
\bibitem [{\citenamefont {{Rokhinson}}\ \emph {et~al.}(2012)\citenamefont
  {{Rokhinson}}, \citenamefont {{Liu}},\ and\ \citenamefont
  {{Furdyna}}}]{Rokhinson}%
  \BibitemOpen
  \bibfield  {author} {\bibinfo {author} {\bibfnamefont {L.~P.}\ \bibnamefont
  {{Rokhinson}}}, \bibinfo {author} {\bibfnamefont {X.}~\bibnamefont {{Liu}}},
  \ and\ \bibinfo {author} {\bibfnamefont {J.~K.}\ \bibnamefont {{Furdyna}}},\
  }\href {http://stacks.iop.org/1367-2630/12/i=6/a=065010} {\bibfield
  {journal} {\bibinfo  {journal} {Nature Physics}\ }\textbf {\bibinfo {volume}
  {8}},\ \bibinfo {pages} {795} (\bibinfo {year} {2012})}\BibitemShut {NoStop}%
\bibitem [{\citenamefont {Niu}\ \emph {et~al.}(2012)\citenamefont {Niu},
  \citenamefont {Chung}, \citenamefont {Hsu}, \citenamefont {Mandal},
  \citenamefont {Raghu},\ and\ \citenamefont {Chakravarty}}]{Chakravarty}%
  \BibitemOpen
  \bibfield  {author} {\bibinfo {author} {\bibfnamefont {Y.}~\bibnamefont
  {Niu}}, \bibinfo {author} {\bibfnamefont {S.~B.}\ \bibnamefont {Chung}},
  \bibinfo {author} {\bibfnamefont {C.-H.}\ \bibnamefont {Hsu}}, \bibinfo
  {author} {\bibfnamefont {I.}~\bibnamefont {Mandal}}, \bibinfo {author}
  {\bibfnamefont {S.}~\bibnamefont {Raghu}}, \ and\ \bibinfo {author}
  {\bibfnamefont {S.}~\bibnamefont {Chakravarty}},\ }\href {\doibase
  10.1103/PhysRevB.85.035110} {\bibfield  {journal} {\bibinfo  {journal} {Phys.
  Rev. B}\ }\textbf {\bibinfo {volume} {85}},\ \bibinfo {pages} {035110}
  (\bibinfo {year} {2012})}\BibitemShut {NoStop}%
\bibitem [{\citenamefont {Tewari}\ and\ \citenamefont {Sau}(2012)}]{Tewari}%
  \BibitemOpen
  \bibfield  {author} {\bibinfo {author} {\bibfnamefont {S.}~\bibnamefont
  {Tewari}}\ and\ \bibinfo {author} {\bibfnamefont {J.~D.}\ \bibnamefont
  {Sau}},\ }\href {\doibase 10.1103/PhysRevLett.109.150408} {\bibfield
  {journal} {\bibinfo  {journal} {Phys. Rev. Lett.}\ }\textbf {\bibinfo
  {volume} {109}},\ \bibinfo {pages} {150408} (\bibinfo {year}
  {2012})}\BibitemShut {NoStop}%
\bibitem [{\citenamefont {Lutchyn}\ \emph
  {et~al.}(2010{\natexlab{a}})\citenamefont {Lutchyn}, \citenamefont {Sau},\
  and\ \citenamefont {Das~Sarma}}]{sau10}%
  \BibitemOpen
  \bibfield  {author} {\bibinfo {author} {\bibfnamefont {R.~M.}\ \bibnamefont
  {Lutchyn}}, \bibinfo {author} {\bibfnamefont {J.~D.}\ \bibnamefont {Sau}}, \
  and\ \bibinfo {author} {\bibfnamefont {S.}~\bibnamefont {Das~Sarma}},\ }\href
  {\doibase 10.1103/PhysRevLett.105.077001} {\bibfield  {journal} {\bibinfo
  {journal} {Phys. Rev. Lett.}\ }\textbf {\bibinfo {volume} {105}},\ \bibinfo
  {pages} {077001} (\bibinfo {year} {2010}{\natexlab{a}})}\BibitemShut
  {NoStop}%
\bibitem [{\citenamefont {Oreg}\ \emph {et~al.}(2010)\citenamefont {Oreg},
  \citenamefont {Refael},\ and\ \citenamefont {von Oppen}}]{oreg10}%
  \BibitemOpen
  \bibfield  {author} {\bibinfo {author} {\bibfnamefont {Y.}~\bibnamefont
  {Oreg}}, \bibinfo {author} {\bibfnamefont {G.}~\bibnamefont {Refael}}, \ and\
  \bibinfo {author} {\bibfnamefont {F.}~\bibnamefont {von Oppen}},\ }\href
  {\doibase 10.1103/PhysRevLett.105.177002} {\bibfield  {journal} {\bibinfo
  {journal} {Phys. Rev. Lett.}\ }\textbf {\bibinfo {volume} {105}},\ \bibinfo
  {pages} {177002} (\bibinfo {year} {2010})}\BibitemShut {NoStop}%
\bibitem [{\citenamefont {Schnyder}\ \emph {et~al.}(2008)\citenamefont
  {Schnyder}, \citenamefont {Ryu}, \citenamefont {Furusaki},\ and\
  \citenamefont {Ludwig}}]{Schnyder08}%
  \BibitemOpen
  \bibfield  {author} {\bibinfo {author} {\bibfnamefont {A.~P.}\ \bibnamefont
  {Schnyder}}, \bibinfo {author} {\bibfnamefont {S.}~\bibnamefont {Ryu}},
  \bibinfo {author} {\bibfnamefont {A.}~\bibnamefont {Furusaki}}, \ and\
  \bibinfo {author} {\bibfnamefont {A.~W.~W.}\ \bibnamefont {Ludwig}},\ }\href
  {\doibase 10.1103/PhysRevB.78.195125} {\bibfield  {journal} {\bibinfo
  {journal} {Phys. Rev. B}\ }\textbf {\bibinfo {volume} {78}},\ \bibinfo
  {pages} {195125} (\bibinfo {year} {2008})}\BibitemShut {NoStop}%
\bibitem [{\citenamefont {{Schnyder}}\ \emph {et~al.}(2009)\citenamefont
  {{Schnyder}}, \citenamefont {{Ryu}}, \citenamefont {{Furusaki}},\ and\
  \citenamefont {{Ludwig}}}]{Schnyder09}%
  \BibitemOpen
  \bibfield  {author} {\bibinfo {author} {\bibfnamefont {A.~P.}\ \bibnamefont
  {{Schnyder}}}, \bibinfo {author} {\bibfnamefont {S.}~\bibnamefont {{Ryu}}},
  \bibinfo {author} {\bibfnamefont {A.}~\bibnamefont {{Furusaki}}}, \ and\
  \bibinfo {author} {\bibfnamefont {A.~W.~W.}\ \bibnamefont {{Ludwig}}},\ }in\
  \href {\doibase 10.1063/1.3149481} {\emph {\bibinfo {booktitle} {Advances in
  Theoretical Pysics: Landau Memorial Conference}}},\ \bibinfo {series}
  {American Institute of Physics Conference Series}, Vol.\ \bibinfo {volume}
  {1134},\ \bibinfo {editor} {edited by\ \bibinfo {editor} {\bibfnamefont
  {V.}~\bibnamefont {{Lebedev}}}\ and\ \bibinfo {editor} {\bibfnamefont
  {M.}~\bibnamefont {{Feigel'Man}}}}\ (\bibinfo  {publisher} {AIP, Melville,
  NY},\ \bibinfo {year} {2009})\ pp.\ \bibinfo {pages} {10--21},\ \Eprint
  {http://arxiv.org/abs/0905.2029} {arXiv:0905.2029 [cond-mat.mes-hall]}
  \BibitemShut {NoStop}%
\bibitem [{\citenamefont {Kitaev}(2009)}]{Kitaev09}%
  \BibitemOpen
  \bibfield  {author} {\bibinfo {author} {\bibfnamefont {A.}~\bibnamefont
  {Kitaev}},\ }in\ \href {\doibase 10.1063/1.3149495} {\emph {\bibinfo
  {booktitle} {Advances in Theoretical Physics: Landau Memorial Conference}}},\
  Vol.\ \bibinfo {volume} {1134},\ \bibinfo {editor} {edited by\ \bibinfo
  {editor} {\bibfnamefont {V.}~\bibnamefont {Lebedev}}\ and\ \bibinfo {editor}
  {\bibfnamefont {M.}~\bibnamefont {Feigel'man}}}\ (\bibinfo  {publisher} {AIP.
  Melville, NY},\ \bibinfo {year} {2009})\ pp.\ \bibinfo {pages} {22--30},\
  \Eprint {http://arxiv.org/abs/0901.2686} {arXiv:0901.2686
  [cond-mat.mes-hall]} \BibitemShut {NoStop}%
\bibitem [{\citenamefont {Tang}\ and\ \citenamefont {Wen}(2012)}]{wen12}%
  \BibitemOpen
  \bibfield  {author} {\bibinfo {author} {\bibfnamefont {E.}~\bibnamefont
  {Tang}}\ and\ \bibinfo {author} {\bibfnamefont {X.-G.}\ \bibnamefont {Wen}},\
  }\href {\doibase 10.1103/PhysRevLett.109.096403} {\bibfield  {journal}
  {\bibinfo  {journal} {Phys. Rev. Lett.}\ }\textbf {\bibinfo {volume} {109}},\
  \bibinfo {pages} {096403} (\bibinfo {year} {2012})}\BibitemShut {NoStop}%
\bibitem [{\citenamefont {Diez}\ \emph {et~al.}(2012)\citenamefont {Diez},
  \citenamefont {Dahlhaus}, \citenamefont {Wimmer},\ and\ \citenamefont
  {Beenakker}}]{Diez12}%
  \BibitemOpen
  \bibfield  {author} {\bibinfo {author} {\bibfnamefont {M.}~\bibnamefont
  {Diez}}, \bibinfo {author} {\bibfnamefont {J.~P.}\ \bibnamefont {Dahlhaus}},
  \bibinfo {author} {\bibfnamefont {M.}~\bibnamefont {Wimmer}}, \ and\ \bibinfo
  {author} {\bibfnamefont {C.~W.~J.}\ \bibnamefont {Beenakker}},\ }\href
  {\doibase 10.1103/PhysRevB.86.094501} {\bibfield  {journal} {\bibinfo
  {journal} {Phys. Rev. B}\ }\textbf {\bibinfo {volume} {86}},\ \bibinfo
  {pages} {094501} (\bibinfo {year} {2012})}\BibitemShut {NoStop}%
\bibitem [{\citenamefont {Ojanen}(2012)}]{Ojanen12}%
  \BibitemOpen
  \bibfield  {author} {\bibinfo {author} {\bibfnamefont {T.}~\bibnamefont
  {Ojanen}},\ }\href@noop {} {\  (\bibinfo {year} {2012})},\ \Eprint
  {http://arxiv.org/abs/1210.3990} {arXiv:1210.3990 [cond-mat.mes-hall]}
  \BibitemShut {NoStop}%
\bibitem [{\citenamefont {Ryu}\ \emph {et~al.}(2010)\citenamefont {Ryu},
  \citenamefont {Schnyder}, \citenamefont {Furusaki},\ and\ \citenamefont
  {Ludwig}}]{Ryu10}%
  \BibitemOpen
  \bibfield  {author} {\bibinfo {author} {\bibfnamefont {S.}~\bibnamefont
  {Ryu}}, \bibinfo {author} {\bibfnamefont {A.~P.}\ \bibnamefont {Schnyder}},
  \bibinfo {author} {\bibfnamefont {A.}~\bibnamefont {Furusaki}}, \ and\
  \bibinfo {author} {\bibfnamefont {A.~W.~W.}\ \bibnamefont {Ludwig}},\ }\href
  {http://stacks.iop.org/1367-2630/12/i=6/a=065010} {\bibfield  {journal}
  {\bibinfo  {journal} {New Journal of Physics}\ }\textbf {\bibinfo {volume}
  {12}},\ \bibinfo {pages} {065010} (\bibinfo {year} {2010})}\BibitemShut
  {NoStop}%
\bibitem [{\citenamefont {Lutchyn}\ \emph
  {et~al.}(2010{\natexlab{b}})\citenamefont {Lutchyn}, \citenamefont {Sau},\
  and\ \citenamefont {Das~Sarma}}]{Lutchyn10}%
  \BibitemOpen
  \bibfield  {author} {\bibinfo {author} {\bibfnamefont {R.~M.}\ \bibnamefont
  {Lutchyn}}, \bibinfo {author} {\bibfnamefont {J.~D.}\ \bibnamefont {Sau}}, \
  and\ \bibinfo {author} {\bibfnamefont {S.}~\bibnamefont {Das~Sarma}},\ }\href
  {\doibase 10.1103/PhysRevLett.105.077001} {\bibfield  {journal} {\bibinfo
  {journal} {Phys. Rev. Lett.}\ }\textbf {\bibinfo {volume} {105}},\ \bibinfo
  {pages} {077001} (\bibinfo {year} {2010}{\natexlab{b}})}\BibitemShut
  {NoStop}%
\bibitem [{\citenamefont {Sticlet}\ \emph {et~al.}(2012)\citenamefont
  {Sticlet}, \citenamefont {Bena},\ and\ \citenamefont {Simon}}]{Sticlet12}%
  \BibitemOpen
  \bibfield  {author} {\bibinfo {author} {\bibfnamefont {D.}~\bibnamefont
  {Sticlet}}, \bibinfo {author} {\bibfnamefont {C.}~\bibnamefont {Bena}}, \
  and\ \bibinfo {author} {\bibfnamefont {P.}~\bibnamefont {Simon}},\ }\href
  {\doibase 10.1103/PhysRevLett.108.096802} {\bibfield  {journal} {\bibinfo
  {journal} {Phys. Rev. Lett.}\ }\textbf {\bibinfo {volume} {108}},\ \bibinfo
  {pages} {096802} (\bibinfo {year} {2012})}\BibitemShut {NoStop}%
\bibitem [{\citenamefont {Chevallier}\ \emph {et~al.}(2012)\citenamefont
  {Chevallier}, \citenamefont {Sticlet}, \citenamefont {Simon},\ and\
  \citenamefont {Bena}}]{Chevalier12}%
  \BibitemOpen
  \bibfield  {author} {\bibinfo {author} {\bibfnamefont {D.}~\bibnamefont
  {Chevallier}}, \bibinfo {author} {\bibfnamefont {D.}~\bibnamefont {Sticlet}},
  \bibinfo {author} {\bibfnamefont {P.}~\bibnamefont {Simon}}, \ and\ \bibinfo
  {author} {\bibfnamefont {C.}~\bibnamefont {Bena}},\ }\href {\doibase
  10.1103/PhysRevB.85.235307} {\bibfield  {journal} {\bibinfo  {journal} {Phys.
  Rev. B}\ }\textbf {\bibinfo {volume} {85}},\ \bibinfo {pages} {235307}
  (\bibinfo {year} {2012})}\BibitemShut {NoStop}%
\bibitem [{\citenamefont {Law}\ \emph {et~al.}(2009)\citenamefont {Law},
  \citenamefont {Lee},\ and\ \citenamefont {Ng}}]{Law09}%
  \BibitemOpen
  \bibfield  {author} {\bibinfo {author} {\bibfnamefont {K.~T.}\ \bibnamefont
  {Law}}, \bibinfo {author} {\bibfnamefont {P.~A.}\ \bibnamefont {Lee}}, \ and\
  \bibinfo {author} {\bibfnamefont {T.~K.}\ \bibnamefont {Ng}},\ }\href
  {\doibase 10.1103/PhysRevLett.103.237001} {\bibfield  {journal} {\bibinfo
  {journal} {Phys. Rev. Lett.}\ }\textbf {\bibinfo {volume} {103}},\ \bibinfo
  {pages} {237001} (\bibinfo {year} {2009})}\BibitemShut {NoStop}%
\bibitem [{\citenamefont {Flensberg}(2010)}]{Flensberg10}%
  \BibitemOpen
  \bibfield  {author} {\bibinfo {author} {\bibfnamefont {K.}~\bibnamefont
  {Flensberg}},\ }\href {\doibase 10.1103/PhysRevB.82.180516} {\bibfield
  {journal} {\bibinfo  {journal} {Phys. Rev. B}\ }\textbf {\bibinfo {volume}
  {82}},\ \bibinfo {pages} {180516} (\bibinfo {year} {2010})}\BibitemShut
  {NoStop}%
\bibitem [{\citenamefont {Romito}\ \emph {et~al.}(2012)\citenamefont {Romito},
  \citenamefont {Alicea}, \citenamefont {Refael},\ and\ \citenamefont {von
  Oppen}}]{Romito}%
  \BibitemOpen
  \bibfield  {author} {\bibinfo {author} {\bibfnamefont {A.}~\bibnamefont
  {Romito}}, \bibinfo {author} {\bibfnamefont {J.}~\bibnamefont {Alicea}},
  \bibinfo {author} {\bibfnamefont {G.}~\bibnamefont {Refael}}, \ and\ \bibinfo
  {author} {\bibfnamefont {F.}~\bibnamefont {von Oppen}},\ }\href {\doibase
  10.1103/PhysRevB.85.020502} {\bibfield  {journal} {\bibinfo  {journal} {Phys.
  Rev. B}\ }\textbf {\bibinfo {volume} {85}},\ \bibinfo {pages} {020502}
  (\bibinfo {year} {2012})}\BibitemShut {NoStop}%
\end{thebibliography}%
\end{document}